\documentclass{emulateapj}

\usepackage[]{amsmath}
\usepackage{amssymb}
\usepackage{natbib}
\usepackage{apjfonts}
\usepackage{graphicx}
\usepackage{multirow}
\usepackage{booktabs}
\usepackage{bm}
\usepackage{xspace}
\usepackage{afterpage}
\usepackage{gensymb}

\shorttitle{Spatially Resolved SFR in cluster galaxies}
\shortauthors{Vulcani et al.}

\usepackage{color}			      
\definecolor{midgray}{gray}{0.4}		
\definecolor{orange}{rgb}{1,0.5,0}    

\usepackage[colorlinks=true,citecolor=midgray,linkcolor=midgray,urlcolor=midgray]{hyperref}        

\newcommand{\HST}{\emph{HST}}

\newcommand{\Ha}{H$\alpha$\xspace} 
\newcommand{\paperxi}{Paper VII\xspace} 
\newcommand{\paperv}{Paper V\xspace}

\begin{document}
\title{THE GRISM LENS-AMPLIFIED SURVEY FROM SPACE (GLASS). VIII. The influence of the cluster properties on  \Ha emitter galaxies at 
0.3$<\lowercase{z}<$0.7} 

\author{Benedetta Vulcani\altaffilmark{1}}
\author{Tommaso Treu\altaffilmark{2}}
\author{Carlo Nipoti\altaffilmark{3}}
\author{Kasper B. Schmidt\altaffilmark{4}}
\author{Alan Dressler\altaffilmark{5}}
\author{Takahiro Morshita\altaffilmark{2,6,7}}
\author{Bianca M. Poggianti\altaffilmark{8}}
\author{Matthew Malkan\altaffilmark{2}}
\author{Austin Hoag\altaffilmark{9}}
\author{Marusa Brada\v{c}\altaffilmark{9}}
\author{Louis Abramson\altaffilmark{2}}
\author{Michele Trenti\altaffilmark{1}}
\author{Laura Pentericci\altaffilmark{10}}
\author{Anja von der Linden\altaffilmark{11}}
\author{Glenn Morris\altaffilmark{12,13}}
\author{Xin Wang\altaffilmark{2}}

\affil{\altaffilmark{1}School of Physics, University of Melbourne, VIC 3010, Australia}
\affil{\altaffilmark{2}Department of Physics and Astronomy, University of California, Los Angeles, CA, USA 90095-1547}
\affil{\altaffilmark{3}Department of Physics and Astronomy, Bologna University, viale Berti-Pichat 6/2, I-40127 Bologna, Italy}
\affil{\altaffilmark{4}Leibniz-Institut für Astrophysik Potsdam (AIP), An der Sternwarte 16, 14482 Potsdam, Germany}
\affil{\altaffilmark{5}The Observatories of the Carnegie Institution for Science, 813 Santa Barbara St., Pasadena, CA 91101, USA}
\affil{\altaffilmark{6} Astronomical Institute, Tohoku University, Aramaki, Aoba, Sendai 980-8578, Japan}
\affil{\altaffilmark{7} Institute for International Advanced Research and Education, Tohoku University, Aramaki, Aoba, Sendai 980-8578, Japan}
\affil{\altaffilmark{8}INAF-Astronomical Observatory of Padova, Italy}
\affil{\altaffilmark{9}Department of Physics, University of California, Davis, CA, 95616, USA}
\affil{\altaffilmark{10}INAF - Osservatorio Astronomico di Roma Via Frascati 33 - 00040 Monte Porzio Catone, I}
\affil{\altaffilmark{11}Stony Brook University Department of Physics and Astronomy Stony Brook, NY 11794}
\affil{\altaffilmark{12}Kavli Institute for Particle Astrophysics and Cosmology, Stanford University, 452 Lomita Mall, Stanford, CA  94305-4085, USA}
\affil{\altaffilmark{13}SLAC National Accelerator Laboratory, 2575 Sand Hill Road, Menlo Park,    CA 94025, USA}

\begin{abstract}
Exploiting the data of the Grism 
Lens-Amplified Survey from Space (GLASS), we characterize  the spatial distribution of star 
formation in 76   high star forming galaxies in 10 clusters at 0.3$<z<$0.7. All these galaxies are likely restricted to first infall.
In a companion paper
we contrast the properties of field and cluster galaxies, whereas here we correlate the properties of \Ha emitters to 
 a number of tracers of the cluster environment  to investigate its role in driving galaxy transformations. 
\Ha emitters are found in the clusters out to 0.5 virial radii, the maximum radius 
covered by GLASS. The peak of the \Ha emission is offset with respect to the peak of  the UV-continuum. 
We decompose this offsets into a radial and tangential component. The radial component 
points away from the cluster center in 60\% of the cases, with 95\% confidence.  The decompositions agree with
cosmological simulations, i.e. the \Ha emission offset correlates with galaxy velocity and ram-pressure stripping signatures. 
Trends between \Ha emitter properties and  surface mass density distributions and X-ray emissions emerge only for unrelaxed clusters. 
 The lack of strong correlations with the global environment does not allow us to identify a unique  environmental effect  originating from
the cluster center. In contrast, correlations between \Ha morphology and local number density emerge. 
We conclude that local effects, uncorrelated to the cluster-centric radius, play a more important role in shaping galaxy properties.

\end{abstract}


\section{Introduction}
Galaxy properties have been found to strongly correlate with environment, at different redshifts 
\citep[e.g.][]{butcher84, dressler80, dressler97, poggianti99, ellis97, lewis02, treu03, gomez03, goto03, postman05,  kauffmann04, grutzbauch11}.
One of the most striking 
differences between galaxies in clusters and in the field is the fraction of star forming galaxies, which
decreases from the densest to the sparsest environments \citep[e.g.][]{vonderlinden10, paccagnella16}. 
The evolution of the star formation activity is paralleled by a corresponding evolution of galaxy morphologies from late- to early-types, whose occurrence is
environment-dependent
\citep{dressler97, fasano00, capak07, poggianti09, oesch10, vulcani11}.

A central question in this picture is
how much  galaxy evolution is driven by
internal processes as opposed to collective phenomena found only in
specific environments. However, as pointed out by \cite{delucia_borg12},  the distinction is not clear
cut: today's clusters correspond to some of the most overdense regions
in the early Universe and therefore we expect their evolution to be
accelerated with respect to average or underdense region, even if
cluster-specific mechanisms were not at all relevant \citep{dressler80, abramson16, lilly16, morishita16}.

Several properties of  dense galaxy clusters give rise to physical processes that have been suggested to transform the 
galaxy morphological and star-forming properties. For example, strong tidal effects  
can distort a galaxy and tear away stars and gas \citep{bekki99}. Rapid, frequent galaxy-galaxy encounters 
induce gravitational perturbations which can greatly affect the stellar and 
gas components of cluster galaxies \citep[also known as harassment,][]{moore96}.
Gas falling onto a cluster is heated by shocks leading to a hot, diffuse intracluster medium (ICM) which permeates the space 
between the galaxies. The ICM can impact the gas within a galaxy 
by either compressing it, leading to triggered star formation \citep{bekkicouch03}, or by removing the 
galaxy gas which is required to fuel star formation and leading to a quenching of 
star formation. This process is known as ram-pressure stripping
\citep{gunngott72}.  Both ram-pressure and tidal stripping by the halo potential can remove the hot gas halo
surrounding the galaxy  \citep[the so-called strangulation,][]{larson80, 
balogh00}.

Disentangling the relative importance of these processes in transforming an infalling galaxy has been the 
subject of much debate. Detailed studies of galaxies affected by cluster 
specific processes are made possible by the signature that each process is expected to leave on the 
spatial distribution of the star formation activity within the galaxy. For example, 
ram pressure is expected to partially or completely strip layers of gas  from a galaxy, leaving a 
recognizable pattern of star formation with truncated \Ha disks smaller than the 
undisturbed stellar disk \citep[e.g.,][]{yagi15}. Strangulation,  depriving the galaxy of its gas reservoir and leaving the 
existing interstellar medium (ISM) in the disk to be consumed by star formation, should instead produce a symmetric pattern. 
Other processes, like strong tidal interactions and mergers, tidal 
effects,  harassment, thermal evaporation \citep{cowie77}, and 
turbulent/viscous stripping \citep{nulsen82} can also deplete the gas in a non-homogeneous way, 
leaving non-symmetric \Ha disks.

Understanding the transformation process has been further complicated by our lack of understanding 
of the impact of cluster growth on  galaxies. Hierarchical cluster
growth occurs via both continuous infall of material from the surrounding filaments and  high impact 
merging of two approximately equal mass clusters. Simulations indicate that a 
significant fraction of both the mass and galaxies in clusters at the current epoch have been accreted 
through minor and major cluster mergers \citep[$\sim$50\%][]{berrier09, 
mcgee09}. Therefore, it is important to understand the impact of this process on the available gas and the galaxies. 
Simulations show that  the high ICM pressure a galaxy experiences during the core-passage phase of a merger can trigger star formation 
\citep{bekki10} while the high relative velocity of ICM and galaxies can enhance ram-pressure stripping 
of the ISM, leading to a sharp truncation of star formation 
\citep{fujita99}. Since the timescales for the star forming phases of galaxies (1-100 Myr) are shorter 
than typical merger timescales ($\sim$Gyrs), a detailed understanding of 
the dynamics and merger 
stage of the cluster are crucial when attempting to interpret the observed galaxy populations.

In this paper we  extend the analysis presented by \citet[hereafter \paperv]{vulcani15} and \citet[hereafter \paperxi]{vulcani16}
and investigate whether cluster properties are able to affect the extent and spatial 
distribution of the \Ha emitters in the  10  Grism 
Lens-Amplified Survey from Space  (GLASS; GO-13459; PI: Treu,\footnote{http://glass.astro.ucla.edu}  \citealt{schmidt14, treu15})
 clusters at 0.3$<z<$0.7. We use resolved spectral information to characterize the gaseous material that has been stripped from 
the galaxy disk by any process. We therefore address how star formation is 
suppressed and look for signs of a dependence of the suppression on cluster morphology.

In \paperv, we illustrated the methodology by focusing on two clusters (MACS0717.5+3745 and 
MACS1423.8 +2404) with different morphologies (one relaxed and one 
merging) and used foreground and background galaxies as a field control sample, for a total of 42 galaxies.  
We investigated trends with  the hot gas density as traced by the X-ray 
emission, and with the surface mass density as inferred from gravitational lens models and found no 
conclusive results. The diversity of morphologies and sizes observed in \Ha\ 
illustrated the complexity of the environmental process that regulate star formation. In \paperxi  
we increased the sample size and used  76 galaxies in clusters and 85 
galaxies in the field to compare the spatial distribution of star formation in galaxies in the two most 
different environments. Here we focus on galaxies in clusters and investigate 
how  the \Ha morphology and the main process thought to be  responsible for the \Ha 
appearance depend on   the clustercentric distance, the hot gas density, the surface 
mass density and the local density. Our goal is to use these sensitive diagnostics
to achieve better insight on the role of the cluster environment in driving galaxy transformations. 

The paper is structured as follows. \S\ref{sec:glass} introduces the dataset and the clusters, \S\ref{sec:sample} presents the  galaxy properties. 
\S\ref{sec:results} presents the main results of this study: we characterize \Ha morphologies as a function of clustercentric distance (\S\ref{sec:dist}) and compare  the observed distribution 
of the projected offsets to cosmological predictions of the orbits of infalling galaxies  (\S\ref{sec:pred}).  We  then characterize \Ha morphologies as a function of
global (\S\ref{sec:global}) and local (\S\ref{sec:local}) cluster properties and the variation of the specific star formation rate (SFR) with environment
(\S\ref{sec:ssfr}).  In \S\ref{sec:conc} we discuss our results and conclude.

We assume $H_{0}=70 \, \rm km \, s^{-1} \, Mpc^{-1}$,
$\Omega_{0}=0.3$, and $\Omega_{\Lambda} =0.7$.  We adopt a  \cite{chabrier03} initial
mass function (IMF) in the mass range 0.1--100
$\textrm{M}_{\odot}$.

\section{The Grism Lens-Amplified Survey from Space}\label{sec:glass}
\subsection{The dataset}

GLASS is a 140-orbit slitless spectroscopic survey with \HST{} in cycle 21. It  has observed
the cores of 10 massive galaxy clusters targeted by the Hubble Frontier 
Fields (HFF; P.I. Lotz, \citealt{lotz16}) and by the Cluster Lensing And Supernova survey 
with Hubble (CLASH; P.I. Postman, \citealt{postman12}) with the  Wide-Field Camera 3 (WFC3) Near Infrared (NIR) grisms G102 
and G141 providing an uninterrupted wavelength coverage from 0.8$\mu$m to 1.7$\mu$m.  
Each cluster was observed at two position angles (PAs) approximately 90 degrees apart to 
facilitate clean extraction of the spectra for objects in the  crowded cluster fields.
The sample of 10 clusters and their properties is presented in in  Table~\ref{tab:clus}.

\begin{table*}[!t]
\caption{Cluster properties \label{tab:clus}}
\centering
\begin{tabular}{llccccccccc} 
\hline
\hline
\bf{cluster} 	&\bf{short}		& \bf{RA}  & \bf{DEC}  & \bf{z} 	&  \bf{phys scale}&\bf{L$_{\rm X}$} 	& \bf{M$_{\rm 500}$}   & \bf{r$_{\rm 500}$}  & 
{\bf PA1} & {\bf PA2}    \\
			& \bf{name}	&\bf(J2000) &(J2000) &				&  (kpc/$^{\prime\prime}$)& (10$^{44}$erg s$^{-1}$)	& \	  (10$^{14}$M$_
\odot$) & (Mpc) & &    \\
\hline
Abell2744       		& A2744 	&00:14:21.2 	&-30:23:50.1 	&0.308  		&4.535	&15.28$\pm$0.39& 17.6$\pm$2.3 &1.65$\pm$0.07 & 135 & 233\\ 
RXJ2248.7-4431  	& RXJ2248 &22:48:44.4 	&-44:31:48.5 	&0.346  	&4.921	&30.81$\pm$1.57& 22.5$\pm$3.3 & 1.76$\pm$0.08& 053 & 133\\
Abell370        		& A370 	&02:39:52.9 	&-01:34:36.5	&0.375      		&5.162	&8.56$\pm$0.37& 11.7$\pm$2.1 &1.40$\pm$0.08 & 155 &253 \\
MACS0416.1-2403 	& MACS0416	& 04:16:08.9 	&-24:04:28.7 	&0.420    &5.532	&8.11$\pm$0.50&  9.1$\pm$2.0 &1.27$\pm$0.09 & 164 &247\\
RXJ1347.5-1145  	& RXJ1347 &13:47:30.6 	&-11:45:10.0  	&0.451 		&5.766	&47.33$\pm$1.2 & 21.7$\pm$3.0 & 1.67$\pm$0.07& 203 & 283\\
MACS1423.8+2404   & MACS1423 	& 14:23:47.8  	& +24:04:40 	& 0.543   	& 6.382	&13.96$\pm$0.52& 6.64$\pm$0.88  & 1.09$\pm$0.05 & 008& 088 \\
MACS1149.6+2223	& MACS1149 	&11:49:36.3 	&+22:23:58.1 	&0.544 	&6.376	&17.25$\pm$0.68& 18.7$\pm$3.0 &1.53$\pm$0.08 & 032 &125\\
MACS0717.5+3745   & MACS0717	& 07:17:31.6  	& +37:45:18 	& 0.546    & 6.400 	&24.99$\pm$0.92& 24.9$\pm$2.7 & 1.69$\pm$0.06 & 020&280 \\
MACS2129.4-0741 	& MACS2129 	&21:29:26.0 	&-07:41:28.0 	&0.589   	&6.524	&13.69$\pm$0.57& 10.6$\pm$1.4 &1.26$\pm$0.05 & 050 &328\\
MACS0744.9+3927  & MACS0744	& 07:44:52.8 	&+39:27:24.0 	&0.686    		&7.087	&18.94$\pm$0.61& 12.5$\pm$1.6 &1.27$\pm$0.05 & 019 & 104\\  
\hline
\end{tabular}
 \tablecomments{J2000 coordinates, redshift, physical scale, X-ray luminosity, M$_{500}$ \citep[from][]{mantz10}, r$_{500}$  and the two position 
angles. }
\end{table*}

Details on the observations and data reduction can be found in \cite{schmidt14, treu15}. 
Briefly, observations follow the dither pattern used for the 3D-HST observations and were 
processed with an updated version of the 3D- HST reduction pipeline\footnote{http://code.google.com/p/threedhst/}
described by \cite{brammer12, momcheva15}. 
All spectra were  visually inspected with the publicly available GLASS inspection GUI, 
GiG\footnote{\href{https://github.com/kasperschmidt/GLASSinspectionGUIs}{github.com/kasperschmidt/GLASSinspectionGUIs}} \citep{treu15},  
in order to  identify and flag erroneous models from the reduction, assess the degree of 
contamination in the spectra and flag and identify strong emission lines and the presence of a continuum. 

As  described in \cite{treu15},  to determine redshifts, templates were compared to each of 
the four available grism spectra independently (G102 and G141 at two PAs 
each) to compute a posterior distribution function for the redshift. 
 Then, with the help 
of the publicly available GLASS inspection GUI for redshifts \citep[GiGz,][]{treu15}, we flagged which grism 
fits are reliable or alternatively entered a redshift by hand if the redshift 
was misidentified by the automatic procedure.
Using GiGz we assigned a quality $Q_z$ to the redshift (4=secure; 3=probable; 2=possible; 1=tentative, but 
likely an artifact; 0=no-$z$). These quality criteria take into account the 
signal to noise ratio of the detection, the possibility that the line is a contaminant, and the identification of 
the feature with a specific emission line. This procedure was carried out 
independently by at least two inspectors per cluster  \citep[see][for 
details]{treu15}.

The full redshift catalogs from the inspection of the 10 GLASS clusters are available at \url{https://archive.stsci.edu/prepds/glass/}.

\subsection{The clusters}\label{sec:clusters}
We make use of all 10 GLASS clusters.
Virial radii $r_{\rm 500}$ have been computed from 
virial masses M$_{\rm 500}$ taken form \cite{mantz10}: 
$$r_{500}=\sqrt[3]{\frac{3}{4\pi}\frac{M_{500}}{500\rho_{cr}}}$$
where 
$\rho_{cr}=\frac{3H^2}{8\pi G}= \frac{3H_0^2}{8\pi G}\times\left[\Omega_\Lambda
+\Omega_0\times \left(1+z)^3\right)\right]$, with G being the gravitational constant = 
$4.29 \times 10^{-9} {\rm (km/s)^2}$  Mpc  ${\rm M}_\odot$.
 
Clustercentric distances have been computed in units of $r_{500}$ from the peak of the X-ray distribution. For 
merging clusters, where more than one peak in the X-ray distribution can be detected, distances have been 
computed from the closest peak. 
Cluster mass maps were produced using the SWUnited reconstruction code described in detail in \citet{bradac04b}
and \citet{bradac09}.  The method uses both  strong and weak lensing mass reconstruction on a non-uniform adapted 
grid. From the set of potential values we determine all 
observables (and mass distribution) using derivatives.  The potential is reconstructed by maximizing the log 
likelihood which uses image positions of multiply imaged sources, weak 
lensing ellipticities, and regularization as constraints. 
Our team has at disposal the cluster mass maps for all clusters, except for MACS0744, which is not ready yet.
The X-ray images are based on Chandra data, and are described in \cite{mantz10} and \cite{vonderlinden14a}.  
For the contours, the images have been adaptively smoothed, 
after removing point sources identified in \cite{ehlert13}. X-ray images are available for all clusters. 
For 
estimating the X-ray emission at the location of the galaxy, we masked the galaxy itself (which can emit in 
X-rays) and computed the average signal in an annulus around the galaxies with inner radius 2$^{\prime\prime}$ and outer radius 5$^{\prime\prime}$.

The local density of a galaxy is defined as the number of its neighbors per unit projected area: $\Sigma=N/A$ in 
number of galaxies per Mpc$^{-2}$.  The projected number densities
 have been estimated from the circular area containing the 5 closest objects: 
 $\Sigma= (5 + 1)/\pi r_5^2$, with $r$ radius of such area \citep[see][]
{morishita16}. Local density estimates are available for only 4 clusters in our sample, 
A2744, MACS0416, MACS0717, and MACS1149, which are the first HFF  clusters with 
complete data.

\section{\Ha maps and Galaxy properties from Paper XI}
\label{sec:sample}
The entire sample and its properties are presented in detail in \paperxi. Briefly, 
from the redshift catalogs, we extract  galaxies with secure
redshift  and consider as cluster members galaxies with
redshift within $\pm$0.03 of the cluster redshift.\footnote{ The limit 0.03 is given at 3$\sigma$, corresponding to a 1-$\sigma$ dispersion of 1000 km/s. This choice is driven by the uncertainty in grism-based redshifts 
owing to limited resolution, of order 0.01.}
Then, we select 
galaxies with visually detected \Ha in emission. We exclude the Brightest Cluster galaxies (BCGs) 
from our analysis, which are not representative of the general cluster galaxy population. 

Overall, our sample includes 76 \Ha-emitting cluster galaxies, distributed among the different clusters as summarized in Table~\ref{tab:gals}. 
 We note that since the GLASS dataset does not typically yield redshifts for cluster passive galaxies (the 4000\AA{} break is too blue for the setup). 

\begin{table}
\caption{Number of galaxies with \Ha in emission in each cluster \label{tab:gals}}
\centering
\begin{tabular}{lcc} 
\hline
\hline
\bf{cluster} 	&\bf{cluster members}  \\
\hline
A2744 	&4\\
RXJ2248 & 3  \\
A370    & 8 \\
MACS0416	&2\\
RXJ1347 &2\\
MACS1423 	&10\\
MACS1149 	&8\\
MACS0717	&16\\
MACS2129 	&8\\
MACS0744	&15\\
\hline
total &76\\
\hline
\end{tabular}
\end{table}

\subsection{Methodology}
\subsubsection{\Ha maps}
Slitless grism observations have high spatial resolution and low spectral resolution, and therefore provide 
images of galaxies in the light of their emission lines for every object in the field of view.
The details of the procedure we followed to make emission line maps of galaxies are described in 
\paperv. Briefly, the \Ha emission line maps are made, separately for 
each PA, by subtracting the continuum from the two-dimensional spectra and masking contaminating flux from nearby objects. 
We then superimposed the \Ha
map onto an image of the galaxy taken with the F475W filter
(rest-frame UV) and onto an image in the F140W (IR). Images are taken from the HFF   \citep{lotz16} or CLASH HST 
\citep{postman12} programs. We use the F475W filter to map relatively recent ($\sim
$100 Myr)  star formation,
and the F140W to trace the older stellar population; 
as opposed to ongoing  ($\sim$10Myr) star formation traced by \Ha. 
Note that for A2744 we used the F435W filter instead, because the F475W 
filter is not available. 

We aligned each map to the continuum 
image of the galaxy, rotating each map by the angle of its PA,
keeping the $y$-offset unaltered with  respect to the continuum. In the dispersion direction, 
there is a degeneracy between the spatial dimension and
the wavelength uncertainty, it is therefore not possible to determine very accurately
the central position of the \Ha map for each PA separately. However, for the cases in
which spectra from both PAs are reliable, which are the vast majority, we used the fact that
the 2 PAs differ by almost $90\degree$, therefore the $x$-direction of
one spectrum roughly corresponds to the $y$-direction of the second
spectrum and vice-versa. We shifted the two spectra
independently along their $x$-direction to maximize the cross correlation between the two maps
to get the intersect.
For the galaxies with reliable spectra in both PAs, we  also measured the real distance between the peak 
of the \Ha emission and the continuum emission, obtained as the quadratic 
sum of the two offsets. 

We also measured the magnitude of the offset between the \Ha and the continuum as projected along the 
cluster radial (off$_r$) and tangential (off$_\theta$) 
directions, determined by  the line connecting the clustercenter and the galaxy center in the continuum light.  
In merging clusters, where more than one cluster center has been 
identified, the closest one to each galaxy  is adopted.  We assigned a positive sign to the radial offset when the peak of the \Ha is 
between the cluster center and 
the peak of the continuum.  

\subsubsection{Additional galaxy properties}\label{sec:gal_prop}
Table 3 in \paperxi summarizes the main galaxy properties that are used also in this analysis. 
Briefly, stellar mass estimates have been  derived  using FAST  v.1.0 \citep{kriek09}  using the 
spectroscopic redshift of each object. CLASH \citep{postman12} or, when 
available, HFF photometry \citep{lotz16} has been adopted. 
For details on stellar mass estimates refer to \cite{morishita16}.

The stellar population properties  have not been fitted for A370, since the final HHF 
observations were not available while this study was carried out.

The surface  SFR density ($\Sigma$\textrm{SFR}, $\textrm{M}_\odot \, \textrm{yr}^{-1}\, \textrm{kpc}^{-2}$) 
and the total SFRs ($\textrm{M}_\odot \, \textrm{yr}^{-1}$),  have been 
derived from \Ha maps. The total SFRs are obtained summing the surface  SFR density  within the Kron 
radius  measured by Sextractor from a combined NIR image  of the 
galaxy. We used the conversion factor derived by \cite{kennicutt94}  and \cite{madau98} 
and corrected SFR estimates for both the scatter due to the [NII] contamination,  applying the locally calibrated correction factor given by 
\cite{james05} and  dust extinction, using the relation given by \cite{garnbest10}. 

As described in \paperxi,  our   $\Sigma
$SFR limit is around $5\times 10^{-2}$ M$_\odot$ yr$^{-1}$ kpc$^{-2}$ for 
SFR$\sim1 \, M_\odot yr^{-1}$ and we use this value as indication of the completeness limit of our sample.

Visual classification of galaxy broad-band morphology in the continuum and of the \Ha line have been performed 
as presented in \paperxi using the publicly available GLASS inspection GUI  for morphologies (GiGm).\footnote{https://github.com/kasperschmidt/GLASSinspectionGUIs} Galaxies have  been 
subdivided into Ellipticals (E), Lenticulars (S0), Spirals (Sp), Mergers (Mer) and 
Irregulars (Irr) and in \Ha regular, \Ha clumpy, \Ha concentrated, \Ha asymmetric/jellyfish. 

We also  attempted to classify the most likely physical processes responsible for altering its continuum and \Ha morphology. 
Five main processes have been proposed: regular, ram pressure, major mergers,   
minor mergers/interaction and other (when none of the above applies).
This is clearly a qualitative and approximate classification scheme, considering that multiple processes might be simultaneously at 
work and that the mapping between morphology and process is not always unique and unambiguous.
As discussed in \paperxi, in spite of the uncertainties, we believe there is merit in categorizing in a self consistent manner the diversity of 
morphological features across environments. In the future, this classification scheme might be replaced with full 2D comparisons with numerical simulations. 
However,  a qualitative visual classification appears to be a useful first step.
 In general, we assigned to the regular class galaxies with regular and undisturbed  \Ha light distribution, 
 to the ram pressure class galaxies where an asymmetry in the \Ha distribution or in the surface brightness is detected.
 We were not able to detect weak cases of ram-pressure stripping, for example 
when a galaxy is at its second or third passage toward the cluster center, but only the strongest ones, when large quantities of gas are still available, 
and the ionized gas is stripped away in a direction that approximately points away from the cluster center. 
 Even though the inspection was not done blindly with respect to the environment,  we did not explicitly take into account the location of the galaxies with respect to the cluster center to characterize this process;
we distinguished between major and minor merger by looking at the same galaxies in the different bands: 
in major mergers the constituents of the mergers are visible both in the F140W and F475W filters, suggesting
they are both massive and luminous; while in minor mergers the F475W filter shows the presence of material 
infalling onto the main galaxies that is not detected in the F140W filter, suggesting that, though 
luminous,  such infalling material is not very massive. 
Examples of the different cases are shown in \paperxi.

\begin{figure*}
\centering
\includegraphics[scale=0.45]{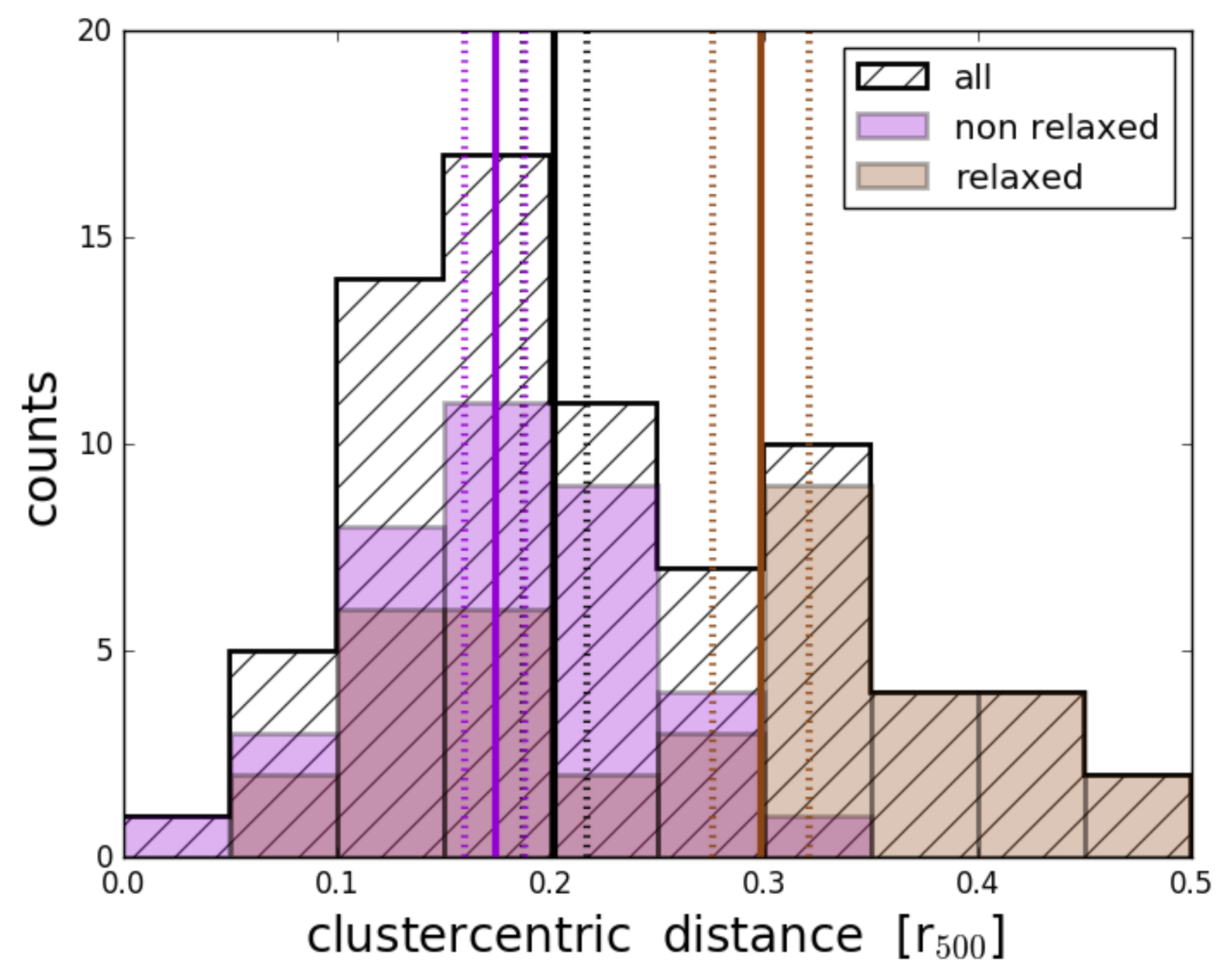}
\includegraphics[scale=0.45]{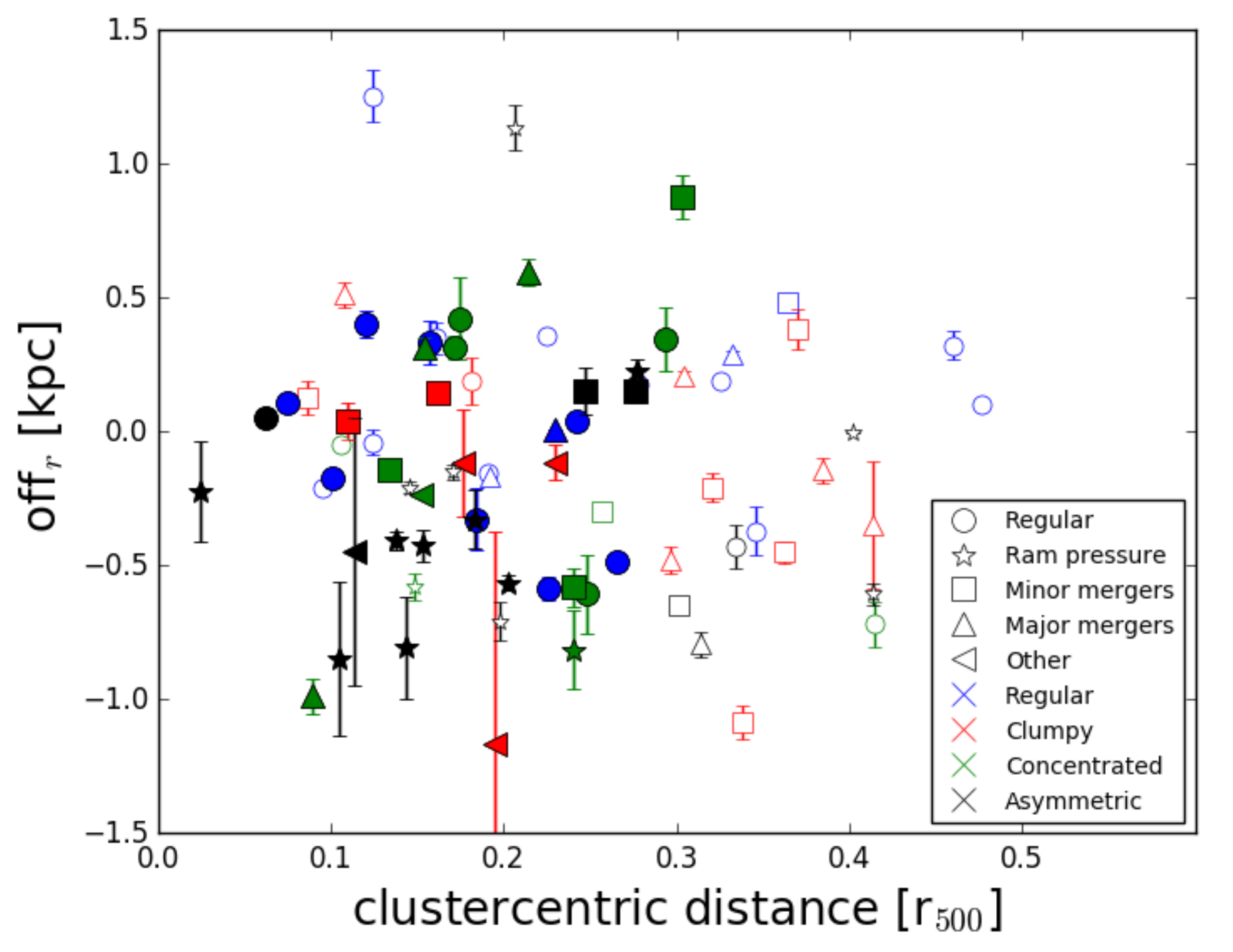}
\includegraphics[scale=0.45]{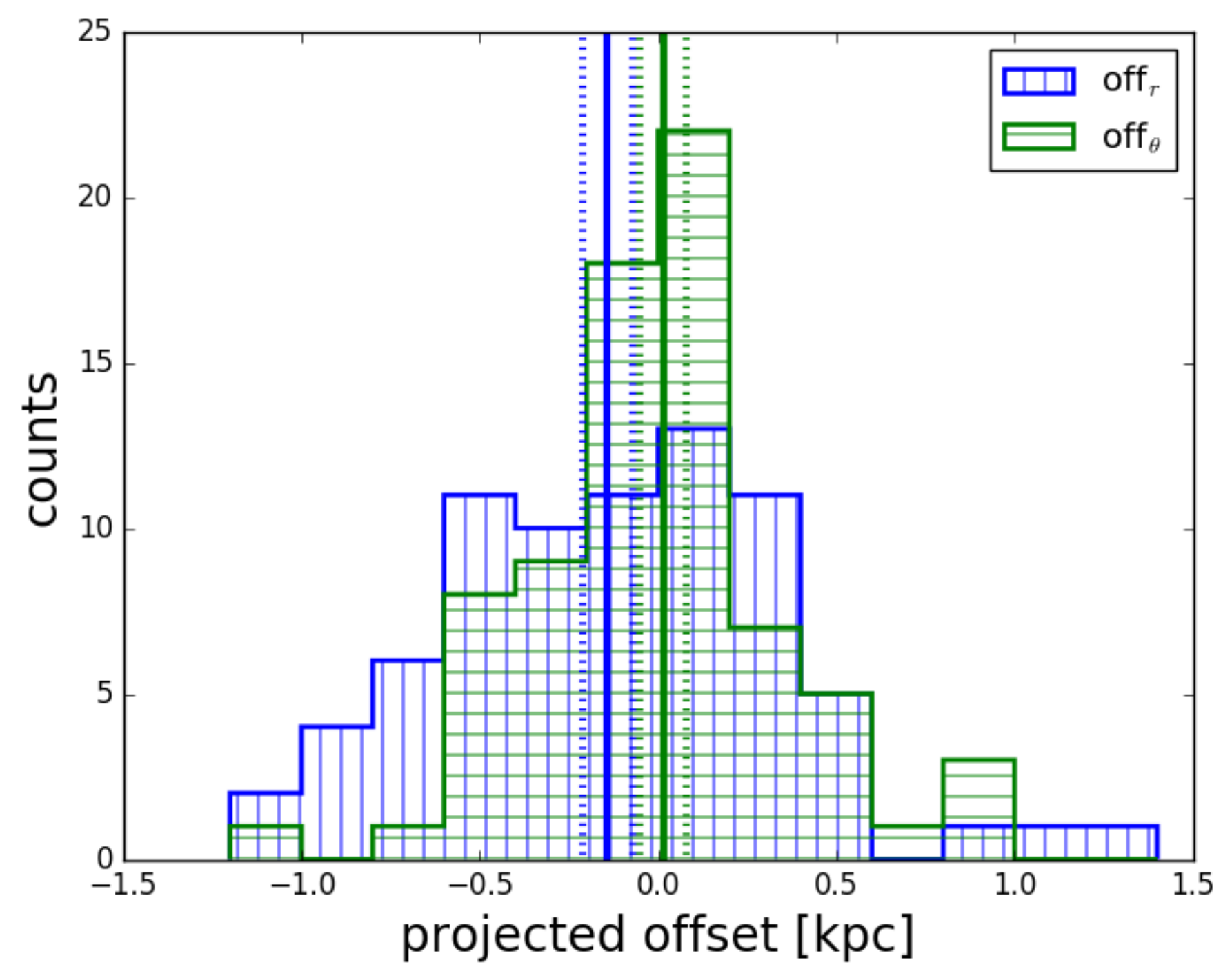}
\includegraphics[scale=0.45]{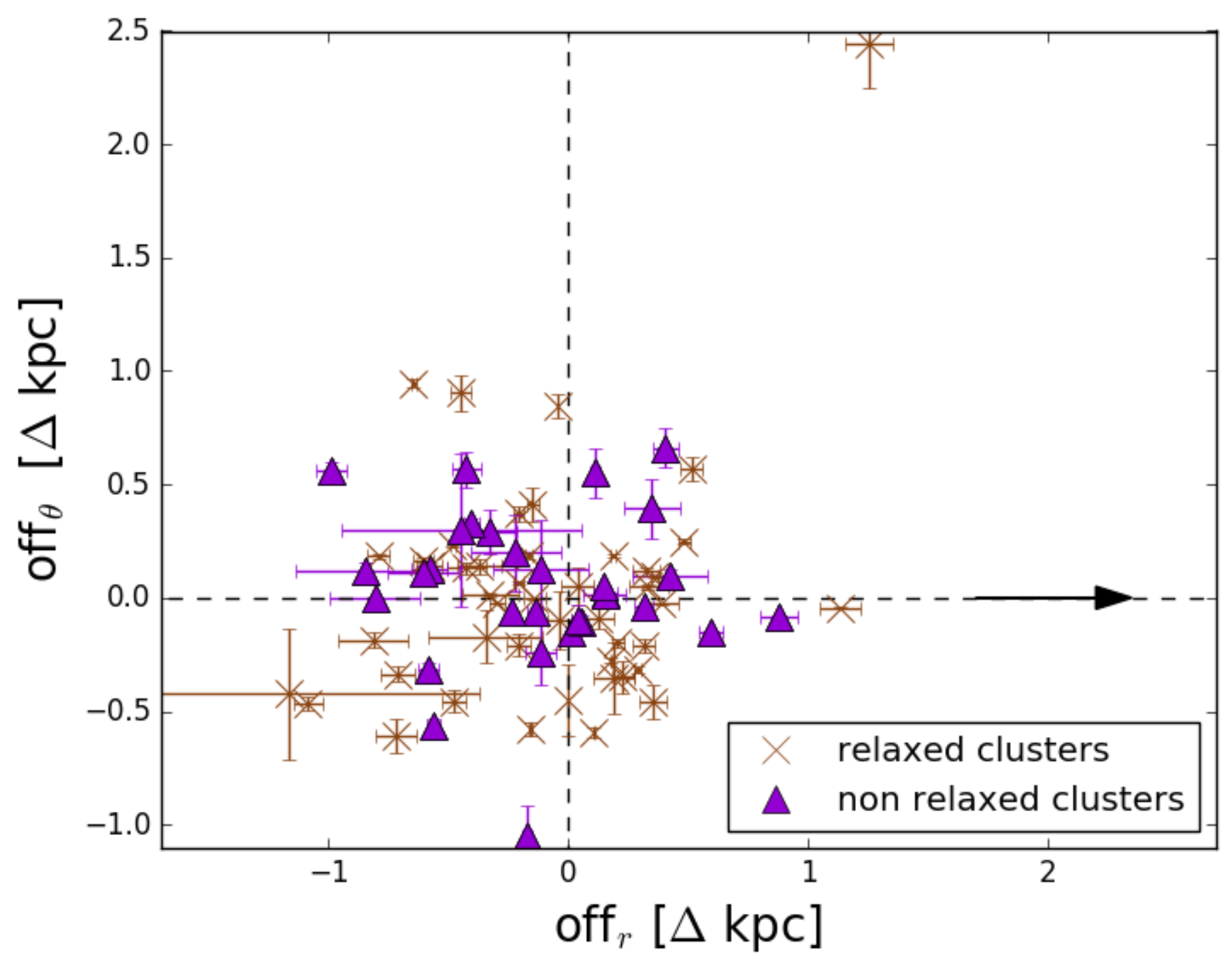}
\caption{{\em Upper left:} Clustercentric distribution of all \Ha emitters.  All galaxies (black), galaxies in unrelaxed clusters (puple) and galaxies in relaxed clusters (brown) are shown. The median value with errors on the median (1.235$\times \sigma/\sqrt{n}$) is also shown. 
{\em Upper right:} Radial projected 
offset (off$_r$) as a function of clustercentric distance, for galaxies with different \Ha morphology (colors) and experiencing different 
physical processes (symbols), as indicated in the label. Galaxies in unrelaxed clusters are shown as filled symbols, galaxies in relaxed clusters as empty symbols.  {\em Bottom left:} radial (blue) and tangential 
(green) projected offsets. Median values along with errors are also shown. 
{\em Bottom right:} Correlation between the tangential and radial projected offset for galaxies in  
relaxed (red crosses) and non relaxed (purple triangles) clusters. The black arrow, located at an arbitrary distance, indicates the 
direction of the cluster centers, the dashed cross the 
UV-continuum light center of the galaxies.  \Ha emitters lie at all distances from the cluster centers, peaking around $r/r_{500}\sim0.2$. Galaxies in unrelaxed clusters are typically closer to the center than galaxies in unrelaxed ones. 
While the  typical tangential offset has a Gaussian distribution peaked at $\Delta \, {\rm off_\theta} =0$, the distribution of radial 
offsets is skewed toward negative values, indicating that \Ha typically points away from the cluster center. The extent of the 
offset does not correlate to clustercentric distances, but there are hints that the offset correlates with some physical processes 
(e.g. ram-pressure stripped galaxies have a more negative offset).  \label{fig:dist}}
\end{figure*}

\section{Results}\label{sec:results}
The focus of the current paper is to correlate the properties of the \Ha emitters to the properties of the clusters in which they are embedded, 
as opposed to \paperxi where we looked at the differences between cluster and field galaxies.
We refer the reader to \paperxi for an exhaustive analysis of the \Ha distribution in the different environments.

\subsection{\Ha morphologies as a function of cluster-centric distance}\label{sec:dist}
We characterize the spatial distribution of the \Ha emitters in terms of cluster-centric distance. 
We note that since the GLASS dataset does not yield redshifts for passive galaxies, we can not characterize
the spatial distribution of all cluster members. 
The upper left panel of Figure~\ref{fig:dist} shows that  galaxies are located within $\sim$0.5r$_{500}$, 
that roughly corresponds to the maximum  coverage of all the 
clusters,  and do not seem to avoid the cluster cores, even though there might be  possible 
projection effects. The distribution peaks around 0.2 r$_{500}$. We distinguish between relaxed (MACS1423, RXJ1347, MACS2129, RXJ2248, MACS0744, for a total of 38 galaxies) and merging/unrelaxed 
(MACS1149, MACS0717, A2744, MACS0416, A370, for a total of 38 galaxies) clusters. 
Since there is no  a unique and clear criterium to distinguish between the two categories, we assume that in unrelaxed clusters more than one X-ray peak is detected, as it will be discussed in Section \ref{sec:global} .   Galaxies in unrelaxed clusters tend to be located closer to the cluster center than galaxies in relaxed clusters. Recall that for merging systems, where more than one peak in the X-ray distribution can be detected, distances have been computed from the closest peak.
 The median value for the former is 0.17$\pm$0.01, that for the latter is 0.30$\pm$0.02. 
 We have also checked 
for mass segregation  and  computed the mean and median galaxy  masses in bins of distance. We found that the typical stellar mass is similar at all distances from the cluster center, suggesting that the mass build-up is not very sensitive to the position of the galaxy in the cluster.

The upper right panel  of Figure~\ref{fig:dist}   quantifies the relation between  the radial 
offset (i.e. the  offset between the peak of the \Ha emission and the peak in the F475W 
filter projected along the  cluster radial direction) and the distance of the galaxy from 
the cluster center. Most of the galaxies have offset within $\pm$0.5 kpc, but there are some 
showing a larger offset.  The typical uncertainty on the offset estimates is $\sim 0.1$ kpc. 
When considering the entire galaxy population as a whole, 
no dependencies on the  cluster-centric distances 
are detected (Spearman correlation=-0.008 with 94\% significance).  Galaxies in relaxed and unrelaxed 
clusters have similar offsets. 
As also  seen in the 
bottom left panel, 60\% of cluster \Ha emitters have negative radial projected 
offset and the distribution is clearly  shifted towards negative values (the median of the 
distribution is -0.14$\pm$0.07 kpc), indicating that for most of the galaxies the \Ha peak 
points away from the cluster center. This finding might suggest that our galaxies are
approaching the cluster center for the first time, and the weakly bound gas is left behind. 
 However, the analysis of the skewness does not support the result: the ratio of the skewness
to the  Standard Error of Skewness (SES)\footnote{The statistical formula for Standard Error of Skewness (SES) for a normal distribution is  
$SES=\sqrt{\frac{6n(n-1)}{(n-2)(n+1)(n+3)}}$.} is 0.23/0.24$\sim$0.83, suggesting that population data are neither positively or negatively skewed.
We will revisit
and try to test quantitatively this hypothesis in the next section. In contrast, the distribution of the  tangential offset  peaks 
around 0 (the median of the distribution is 0.01$\pm$0.07 kpc, skewness/SES =0.15/0.24$\sim$0.62), indicating no preferential 
direction. A K-S test confirms that the two distributions are different (i.e. 4\% probability of being drawn from the same parent
distribution). If we consider only the 39/76 galaxies for which
we have two orthogonal  spectra and therefore the offset is better constrained, we find the same trends, indicating our results
are robust against uncertainties. 
No strong differences are found for galaxies in relaxed and unrelaxed clusters. 

The bottom right panel of Figure~\ref{fig:dist} correlates the tangential to the radial 
offset. As already noticed, there is no preferential direction for the tangential offset, while 
the radial offset is directed away from the cluster center. 
No differences emerge for relaxed and unrelaxed clusters

Galaxies with different \Ha morphologies and experiencing different physical processes 
are highlighted in the upper right panel of Figure~\ref{fig:dist}. 
Ram-pressure stripped galaxies with asymmetric morphology are preferentially found between 0.1 
and 0.3 $r_{500}$, and tend to have negative radial offset, indicating that in 
these galaxies the \Ha distribution is strongly influenced and shows a systematically different 
distribution than the existing stellar population.  In contrast, galaxies of the other 
types are not clustered.

\subsubsection{Comparison of galaxy infall to cosmological simulations} \label{sec:pred}
In the previous Section we have found that the magnitude of the offset
might give us an indication on the process operating on galaxies. In
addition, it might also carry information about the orbit along which
a galaxy is traveling through the ICM. In particular, if the offset is
due to ram pressure, its direction  is expected to trace
the direction of the galaxy velocity.  Therefore, the ratio ${\rm
  off}_r/|{\rm off}_{\theta}|$ between the radial and the tangential
offsets can be taken as a proxy for the ratio $v_r/|v_{\theta}|$
between the radial and the one-dimensional tangential components of the
galaxy velocity at the time of the observation ($v_r$ and $v_\theta$
are two of the three components of the velocity vector in spherical
coordinates). As ${\rm off}_r$ is defined so that it is positive when
the peak of the \Ha emission is closer to the cluster centre than the
continuum emission, we expect that an infalling galaxy ($v_r<0$) has
negative ${\rm off}_r$.  In this Section we compare our observed
radial and tangential offsets to the cosmological predictions of the
orbits of infalling satellites onto galaxy clusters. For simplicity in
what follows we just identify ${\rm off}_r/|{\rm off}_{\theta}|$ with
$v_r/|v_{\theta}|$, neglecting all possible sources of difference
between the two quantities (for instance the offset ratio is a
projected quantity, while the velocity ratio is an intrinsic
quantity). We note that the proxy is an underestimate of the real
offset, since we would not measure any offset for objects which are infalling along the line of sight, yet they would have large $v_r/v_\theta$.

As reference for the cosmological predictions we take the results of
\citet{Jia15}, who studied the distribution of the orbital parameters
of infalling satellite halos  in a $\Lambda$ cold dark matter
($\Lambda$CDM) cosmological $N$-body simulation. In particular
\citet{Jia15} provide the distributions of $V/V_{200}$
and $V_r/V$ as functions of host-halo mass and
satellite-to-host halo mass ratio, where $V$ is the satellite's speed at $r_{200}$, $V_r$ is
the radial component of the satellite's velocity at $r_{200}$ and
$V_{200}$ is the host-halo circular velocity at $r_{200}$.
\citet{Jia15} parameterize the distribution of $V/V_{200}$ with the
three dimensionless parameters $\gamma$, $\sigma$ and $\mu$, and the
distribution of $V_r/V$ with the dimensionless parameter $B$ (see
section 3.4 in that paper). Here we fix $\gamma=0.05$, $\sigma=0.118$,
 $\mu=1.236$, and $B=3.396$, which are the best-fitting values for
host-halo mass $10^{14}M_{\odot}$ and satellite-to-host mass ratio
0.05-0.005 from \citet{Jia15} (note, however, that our results are not strongly dependent
on this specific choice).

Assuming that the host halo is spherical and exploiting the fact that
energy and angular momentum are conserved (neglecting tidal stripping
and dynamical friction), for each orbit of given $V/V_{200}$ and
$V_r/V$ it is straightforward to compute
the ratio $v_r/|v_\theta|$, 
at each radius $r<r_{200}$. Specifically, we assumed that the host halo has a NFW
\citep{Nav96} density distribution with concentration $c_{200}=4$ (in
this case $r_{500}/r_{200}\simeq 0.65$).

\begin{figure}
\centering
\includegraphics[scale=0.4]{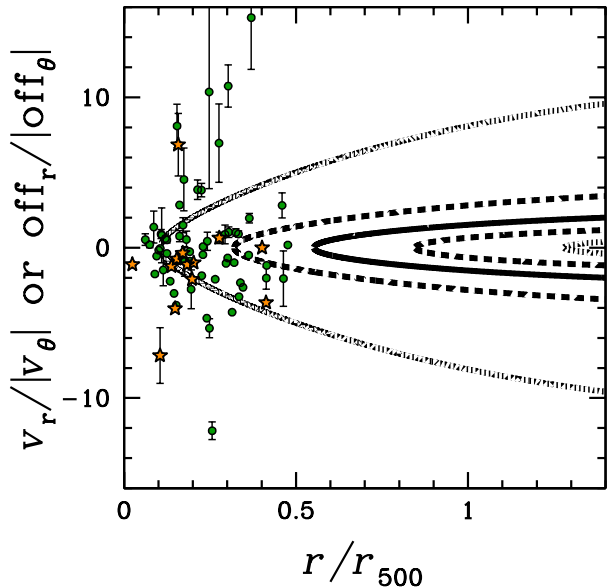}
\caption{
 Observed projected
  radial-to-tangential offset ratio ${\rm off}_r/|{\rm off}_{\theta}|$
  as a function of distance from the cluster centre. 
  The stars represent the galaxies belonging to the ram-pressure
  class, while the other galaxies are represented by circles. Note
  that four galaxies are not shown 
  because they have
  $|{\rm off}_r/{\rm off}_{\theta}|>16$. 
  Radial-to-tangential velocity ratio $v_r/|v_{\theta}|$ as a function
  of radius for characteristic orbits of the cosmological satellite
  orbit distribution estimated by \citet{Jia15} are shown as the various curves.  The orbits
  correspond to the following percentiles of the predicted
  distribution of $V_r/V$ (where $V$ is the speed and $V_r$ is the
  radial component of the velocity at $r_{200}$): 50th (solid curve),
  25th and 75th (dashed curves), and 5th and 95th (dotted curves).  
  The observed sample of cluster galaxies  traces only the $\sim$25\% most radial orbits of the
  cosmological distribution.}
\label{fig:velprofile}
\end{figure}

As a first comparison between the cosmological predictions and the
observations, we look at the behavior of the offset and velocity
ratios as functions of distance from the cluster centre (for
simplicity here we identify the projected observed cluster distance
with the intrinsic orbital radius $r$). In Figure~\ref{fig:velprofile}
the radial distribution of the observed offset ratios is compared with
a few orbits characteristic of the cosmological orbit distribution.
Since \citet{Jia15} find that the distribution of $V/V_{200}$ is
relatively narrow, for simplicity, to compute the theoretical curves
in Figure~\ref{fig:velprofile}, we fix $V/V_{200}=1.236$ (the average
value of the best fit of the distribution) and we sample the
distribution in $V_r/V$ by selecting orbits corresponding to the 5th,
25th, 50th, 75th and 95th percentiles. Figure~\ref{fig:velprofile}
shows, as expected, that the observed sample (confined within
$r/r_{500}<0.5$) traces only the $\approx 25\%$ most radial orbits of
the cosmological distribution: the bulk of the cosmological orbits do
not plunge deep enough into the cluster potential \citep[see
  also][]{crane86}. One caveat is that in the analysis above we have
neglected dynamical friction. However, this effect is negligible, at
least for the first pericentric passage,  as we verified by
  running $N$-body simulations for all the orbits represented in
  Figure~\ref{fig:velprofile}. In these simulations, run with the
  collisionless $N$-body code {\sc fvfps} \citep[][]{Lon03,Nip03}, we
  have followed, starting from $r_{500}$, the orbit of an infalling
  galaxy, represented as a particle with mass $0.005M_{200}$, in an
  isotropic NFW halo with concentration $c_{200}=4$, realized with
  $N\simeq 10^6$ particles ($M_{200}$ is the total mass of the halo,
  which is truncated exponentially at $r_{200}$). The set-up and
  technical characteristics of these simulations are identical to
  those described by \citet{Nip17}.

\begin{figure}
\centering
\includegraphics[scale=0.8]{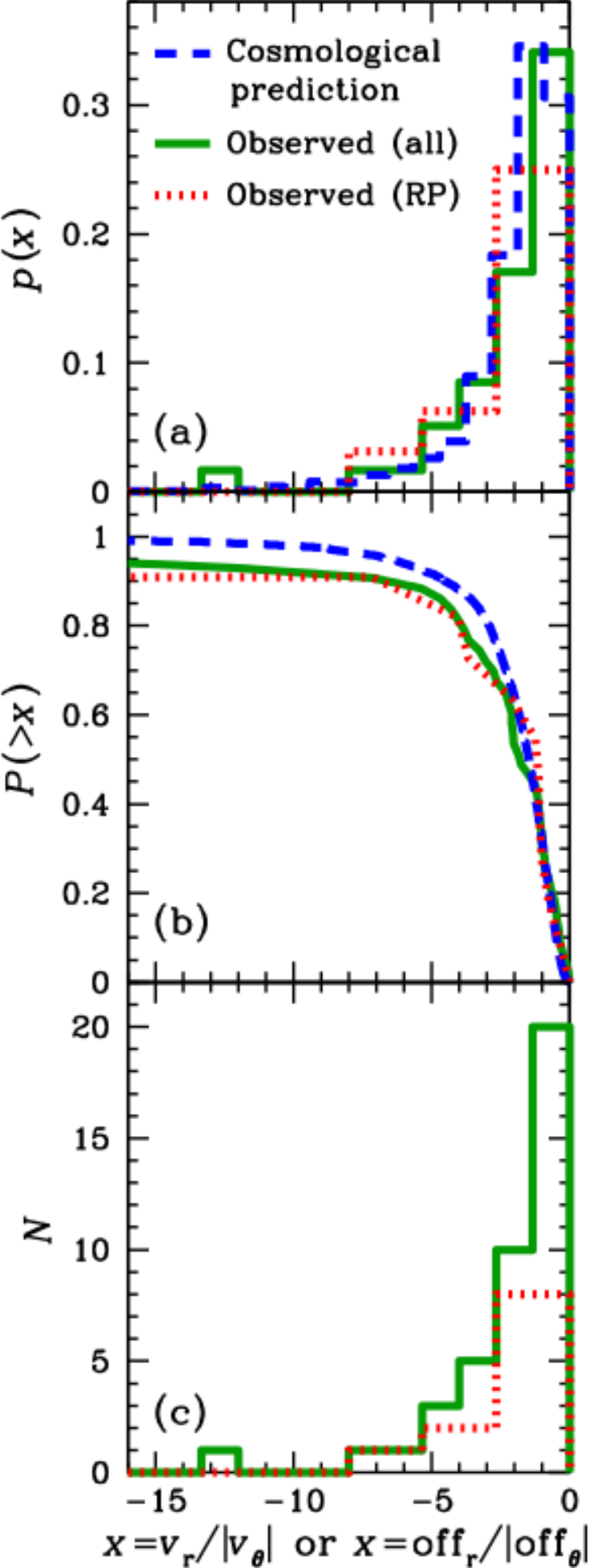}
\caption{ {\em Panel a.} Probability distribution ($p$) of the observed
  radial-to-tangential offset ratio ${\rm off}_r/|{\rm off}_{\theta}|$
  for the 44 galaxies in our sample with negative ${\rm off}_r$ (solid
  curve; sample ``all'') and for the sub-sample of 12 galaxies
  belonging to the ram-pressure class (dotted curve; sample ``RP'').
  The dashed curve represents the probability distribution of the
  radial-to-tangential velocity ratio $v_r/|v_{\theta}|$ for the mock
  sample of infalling ($v_r<0$) galaxies with the same radial
  distribution as sample ``all''. At fixed radius, the values of $v_r$
  and $|v_{\theta}|$ are generated by following orbits that at
  $r_{200}$ have the orbital parameter distribution estimated by
  \citet{Jia15} from a $\Lambda$CDM $N$-body simulation. The
  distribution of $v_r/|v_{\theta}|$ of the mock galaxies of the
  sample ``RP'', not shown, is almost indistinguishable from that of
  the mock galaxies of sample ``all''.  {\em Panel b.} Cumulative
  distributions ($P$) of the two observed samples and of the
  cosmological prediction. {\em Panel c.} Number of galaxies per bin
  of ${\rm off}_r/|{\rm off}_{\theta}|$ for the two observed samples
  (the bins are those used in panel a).}
\label{fig:cumul}
\end{figure}

Figure~\ref{fig:velprofile} suggests that the observed \Ha cluster
galaxies might represent the radial-orbit selected tail of the distribution
of cosmological satellites.  If the cosmological prediction is
correct, and if ${\rm off}_r/|{\rm off}_{\theta}|$ is a proxy for
$v_r/|v_\theta|$, at each radius ${\rm off}_r/|{\rm off}_{\theta}|$
should be distributed as the predicted $v_r/|v_\theta|$. To verify
whether this is actually the case, we select, among the observed
galaxies only the subsample (44 galaxies) with ${\rm off}_r<0$, that,
under our hypothesis, are infalling galaxies, for which the mapping
between ${\rm off}_r/|{\rm off}_{\theta}|$ and $v_r/|v_\theta|$ should
be more justified. In principle ${\rm off}_r/|{\rm
    off}_{\theta}|$ might be a proxy for $v_r/|v_\theta|$ also for
  galaxies that are receding from the centre of the cluster ($v_r>0$),
  but  this model is too simple to
  describe a system that has already passed the pericenter.  For
comparison with this subsample we generate a sample of 4400 mock
galaxies with the same radial distribution. 
The orbital parameters of these mock galaxies are extracted from the
distributions of $V/V_{200}$ and $V_r/V$ given by \citet{Jia15} with
the values of the parameters reported above.  The mock sample of
galaxies can be used to numerically compute the distribution of
$v_r/|v_\theta|$ at each observed radius to be compared with the
observed values of ${\rm off}_r/|{\rm off}_{\theta}|$.  In order to
verify whether the observed and mock samples are consistent, we first
compute the probability distribution of ${\rm off}_r/|{\rm
  off}_{\theta}|$ for the 44 observed galaxies and the probability
distribution of $v_r/|v_\theta|$ for the 4400 mock galaxies. From
Figure~\ref{fig:cumul}a, where these distributions are plotted, it is
apparent that there is qualitative agreement between the observed and
theoretical histograms. When the cumulative distributions are
considered (inset in Figure~\ref{fig:cumul}b), there appears to be a
discrepancy for large values $|{\rm off}_r/{\rm off}_{\theta}|$ and
$|v_r/v_\theta|$ ($x \lesssim -5$), but this discrepancy is not
statistically significant, because the number of observed galaxies in
this tail of the distribution is small (see
Figure~\ref{fig:cumul}c). This can be quantified with a K-S test,
which gives a probability of $51\%$ that the two samples are extracted
from the same parent population. As a further statistical test, for
each galaxy of the sample we computed how it ranks within the
distribution of mock galaxies at the same radius. If the distributions
are consistent the quantiles must be distributed uniformly. According
to the K-S test, the probability that the quantiles are extracted from
a uniform distribution is 25\%.

We repeated the above analysis for the subsample of 12 galaxies with
${\rm off}_r<0$ that we visually classified as affected by
ram-pressure stripping (see Sec. \ref{sec:gal_prop}, stars in
Figure~\ref{fig:velprofile}), for which our model is expected to work
best (in this case we created a mock sample of 1200 galaxies). We find
again that the cumulative distribution of the offset ratios
(Figure~~\ref{fig:cumul}b) is consistent with the theoretical
expectation, as supported by the K-S test, which gives a probability
of 71\% that the two samples (mock and observed) are extracted from
the same parent population.  In this case the probability that the
distribution of the quantiles is uniform is 48\%. 

Bearing in mind the small sample size, we conclude that the observed
distribution of offset ratios for the infalling galaxies is consistent
with the cosmological predictions.  This finding is even more
significant when we consider only the infalling galaxies we labeled as
affected by ram-pressure stripping, therefore cosmological predictions
support our classification scheme.


\begin{figure*}[!t]
\centering
\includegraphics[scale=0.346]{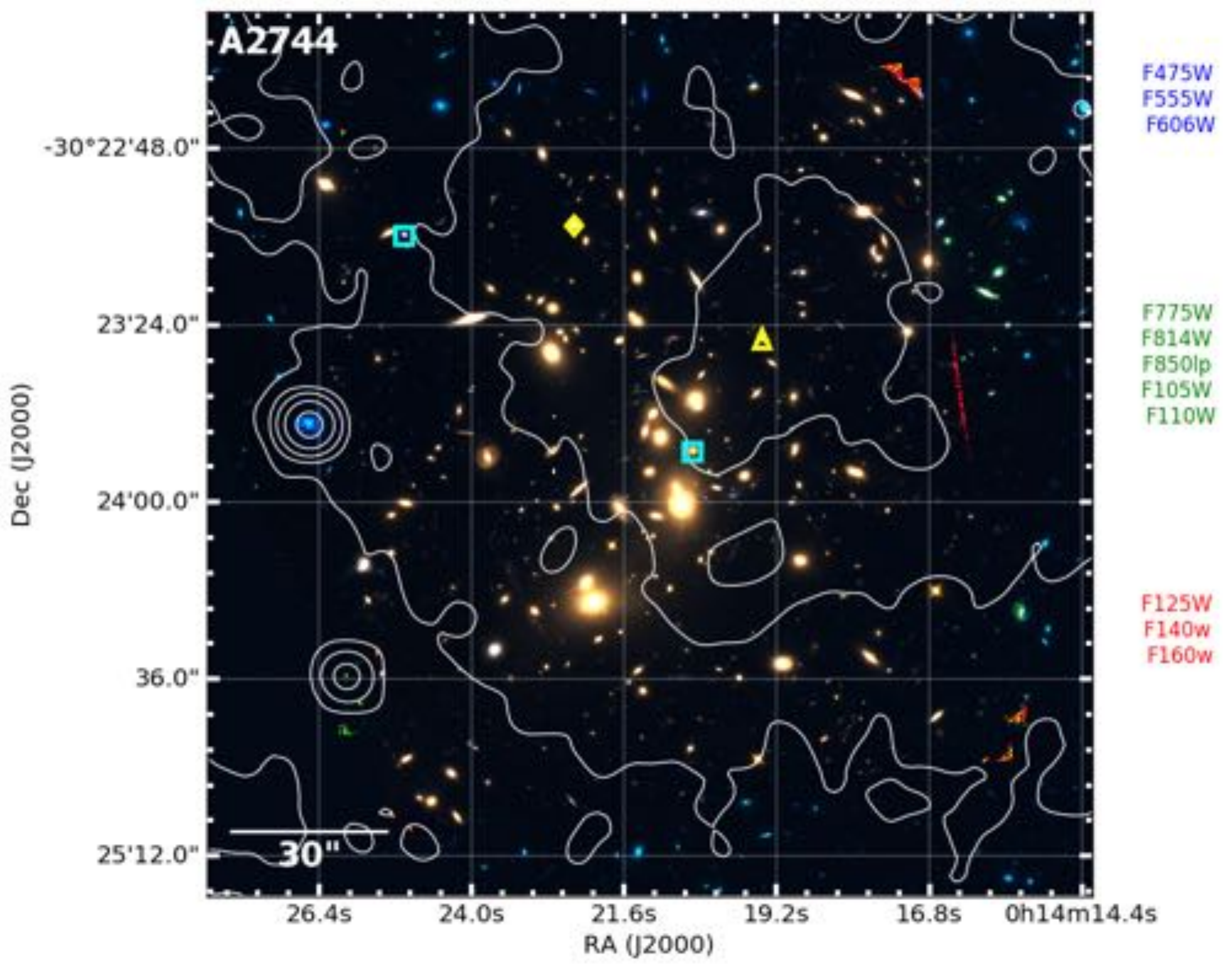}
\includegraphics[scale=0.346]{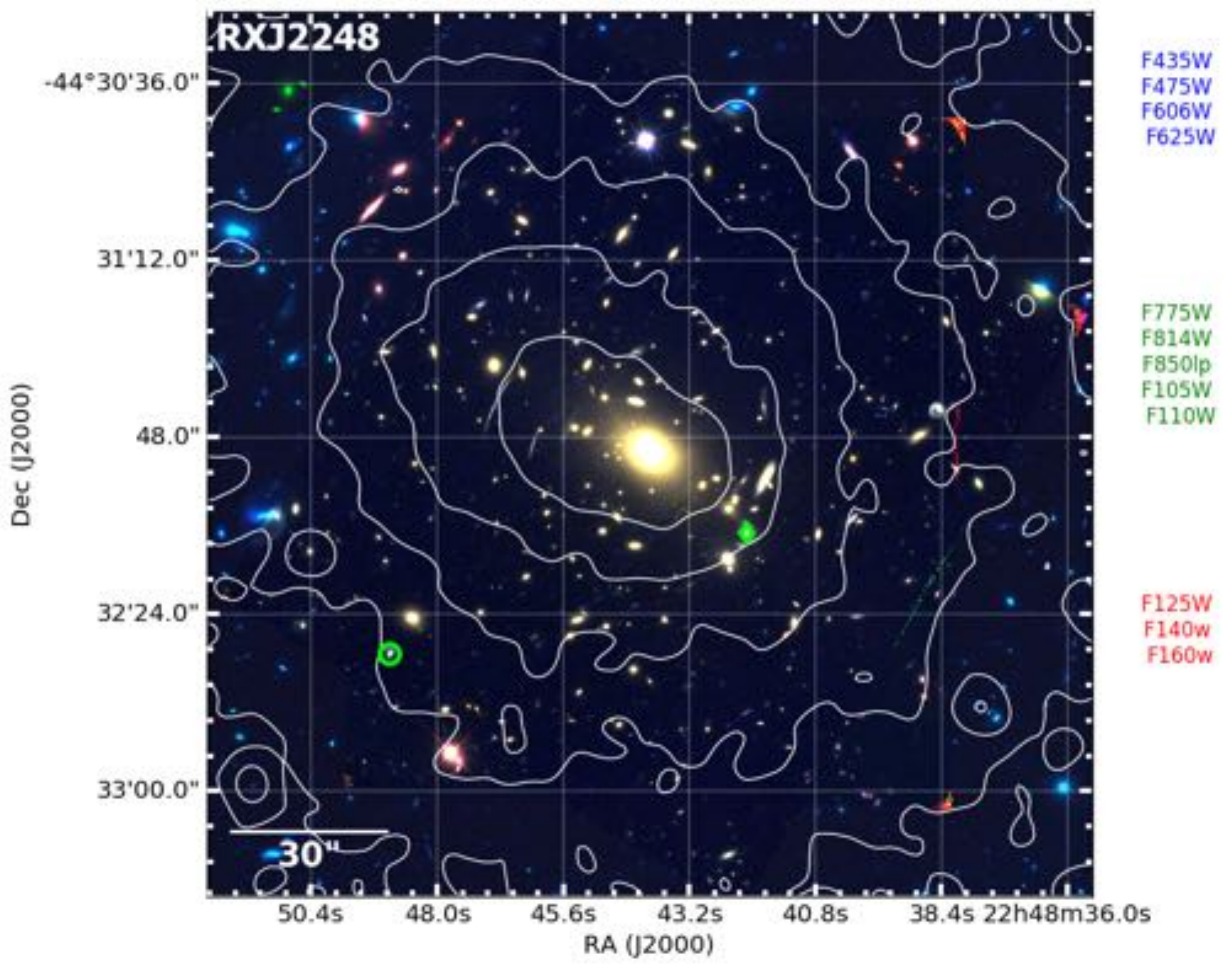}
\includegraphics[scale=0.346]{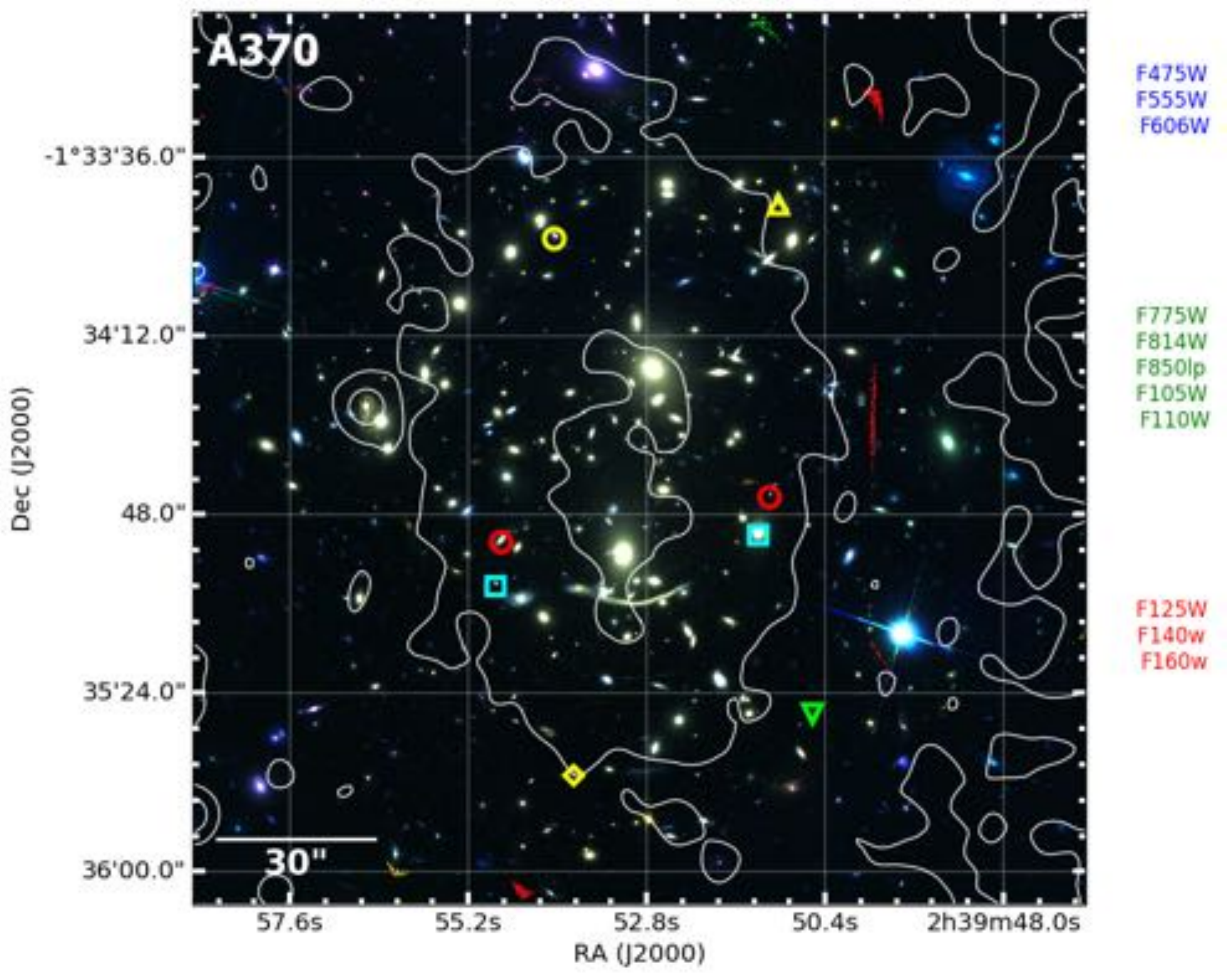}
\includegraphics[scale=0.346]{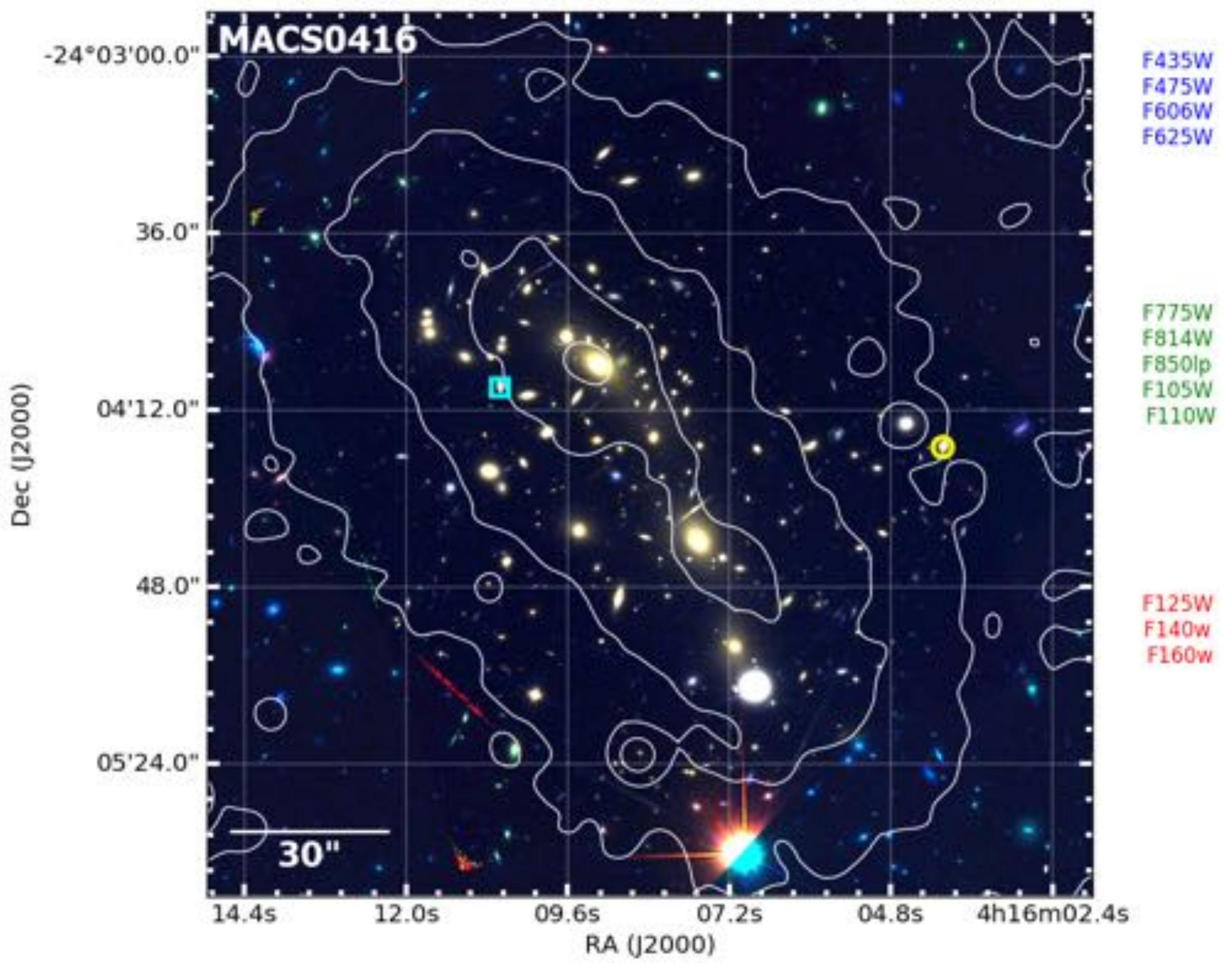}
\includegraphics[scale=0.346]{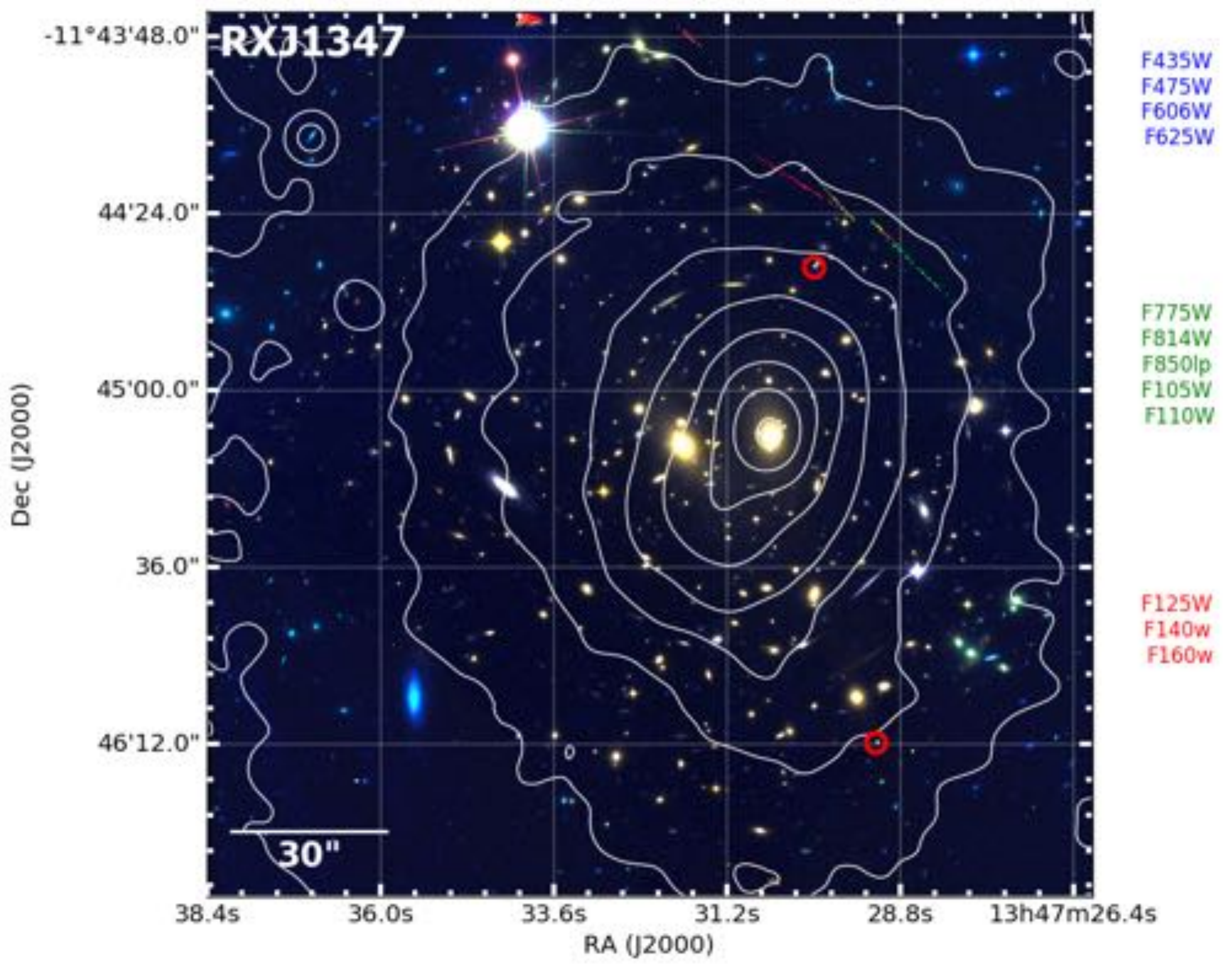}
\includegraphics[scale=0.346]{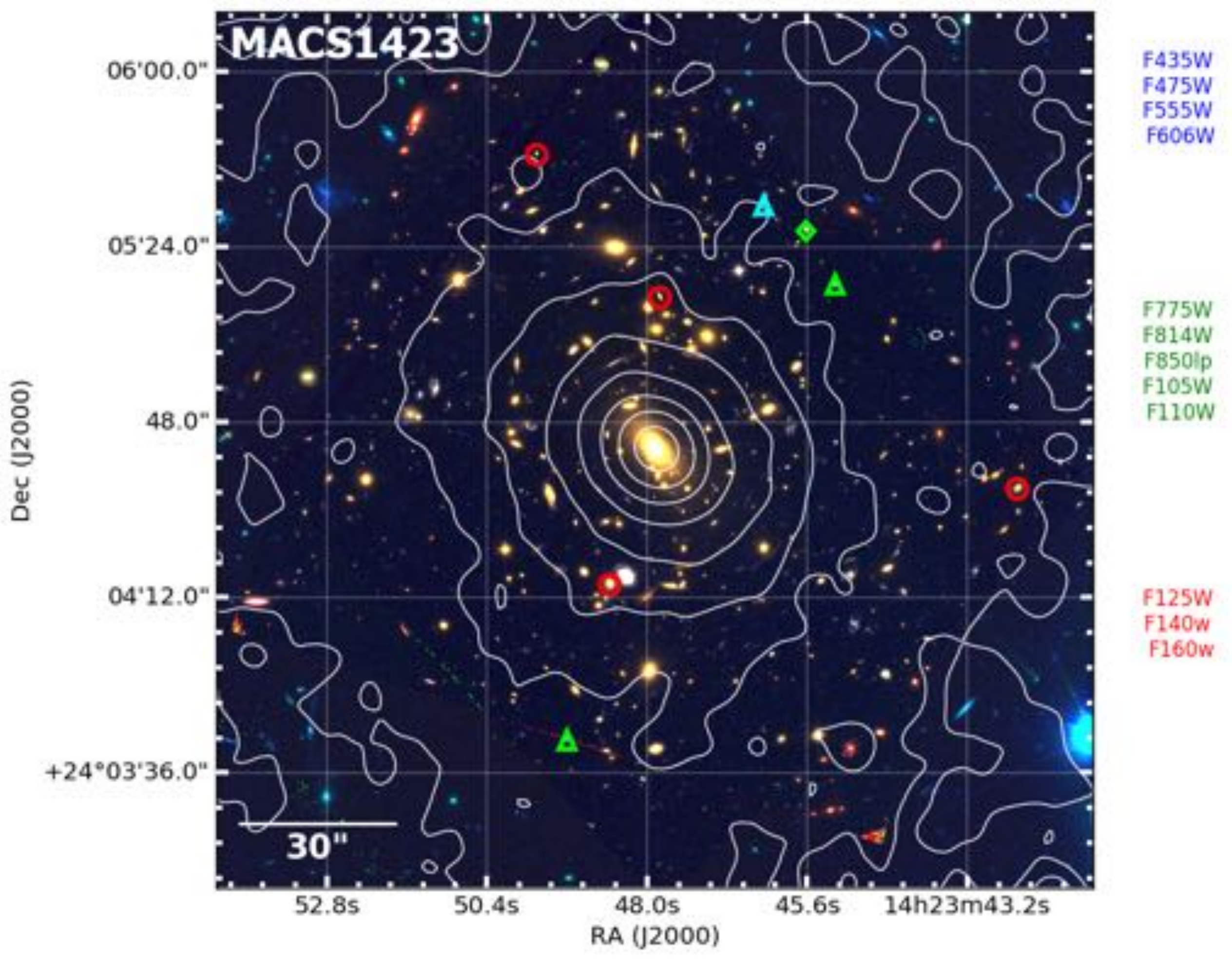}
\caption{Color composite image of  the 10 GLASS clusters. Images are based on the CLASH \citep{postman12} or HFF \citep{lotz16} HST data. The blue, green, and red channels are composed 
by the filters on the right.  X-ray count contours are  overplotted.  Contours are  spaced on a log scale from  0 to 1 counts/s/kpc$^2$. \Ha emitters  with different \Ha morphologies (different colors)  and experiencing different 
processes (different symbols) are also highlighted. Red symbols: regular \Ha, green 
symbols: clumpy \Ha, yellow symbols: concentrated \Ha, cyan symbols: asymmetric \Ha. 
Circles: regular process, squares: ram-pressure stripping, triangles: major-mergers, 
diamonds: minor mergers, inverted triangles: other.  
\label{xray}}
\end{figure*}
\subsection{\Ha morphologies as a function of hot gas density and surface mass density}\label{sec:global}

\renewcommand{\thefigure}{\arabic{figure} (Continues)}
\addtocounter{figure}{-1}

\begin{figure*}[!t]
\centering
\includegraphics[scale=0.346]{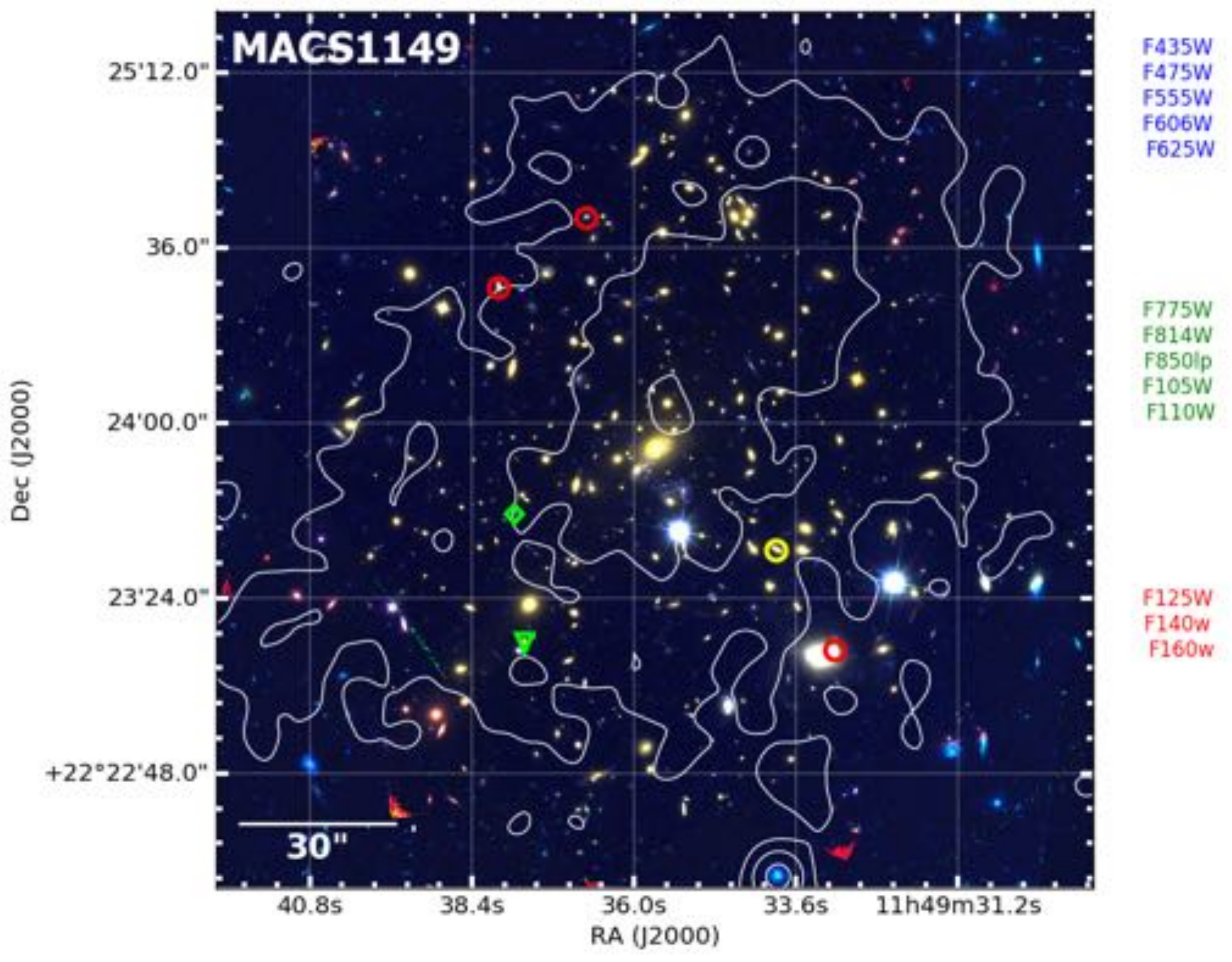}
\includegraphics[scale=0.346]{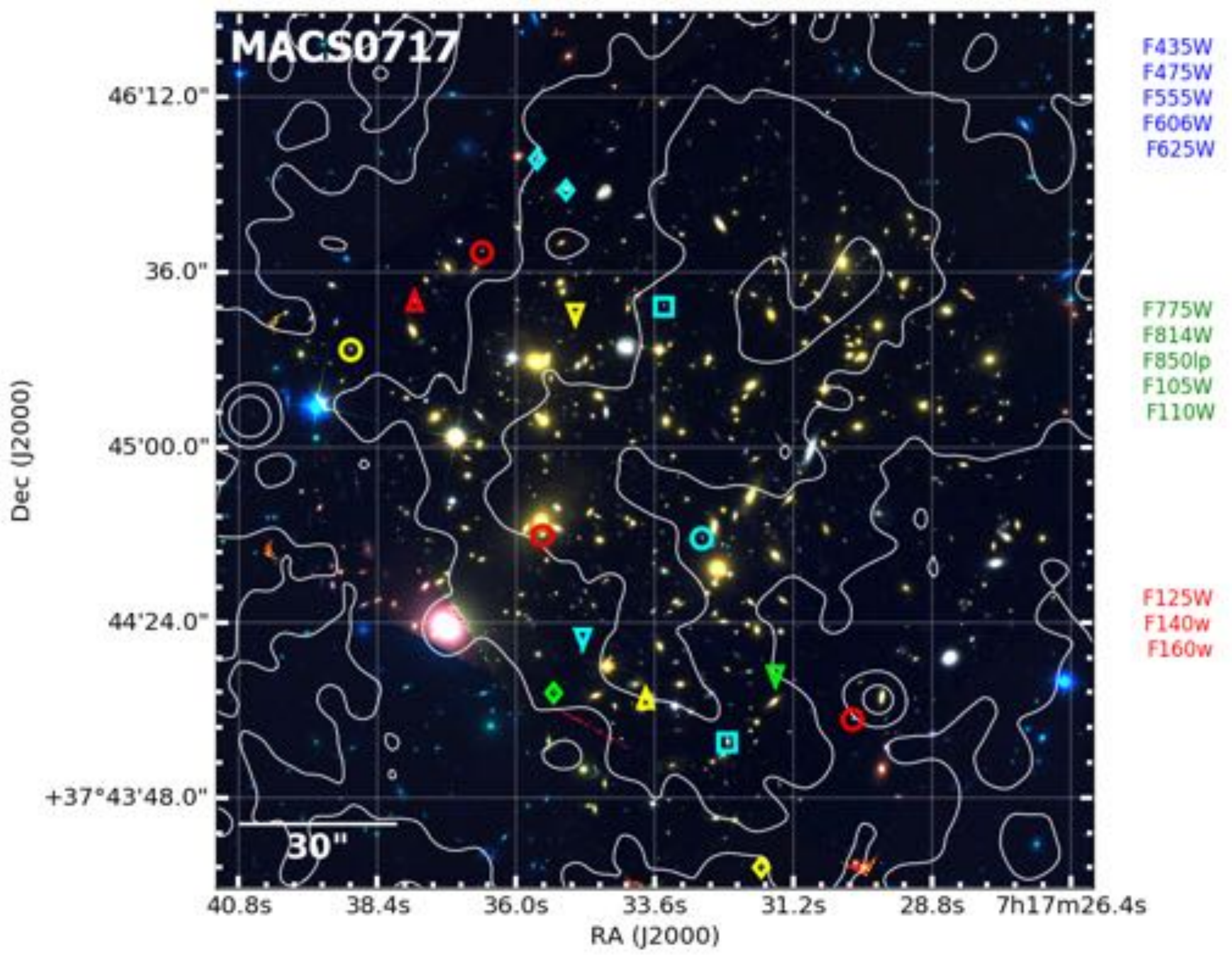}
\includegraphics[scale=0.346]{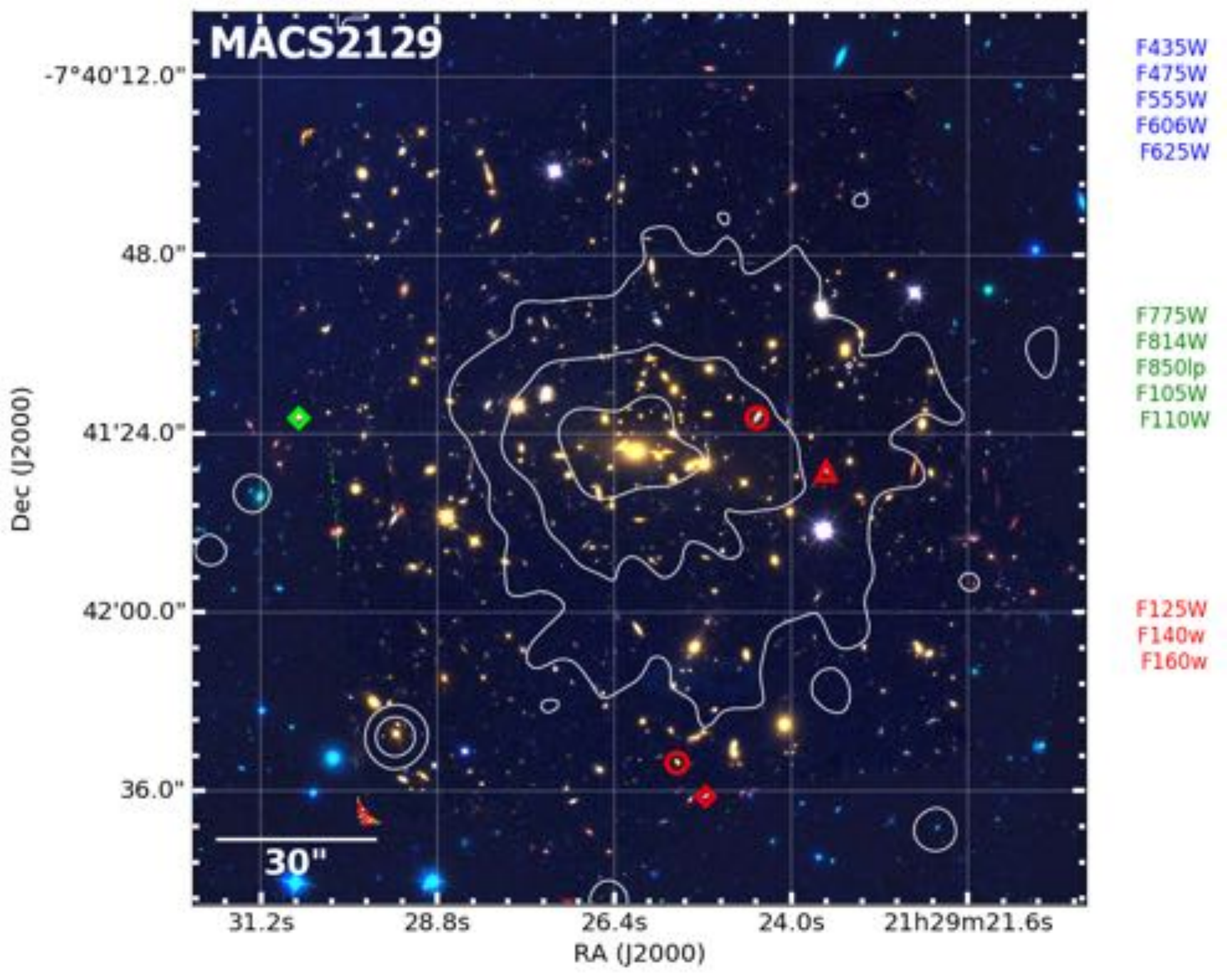}
\includegraphics[scale=0.346]{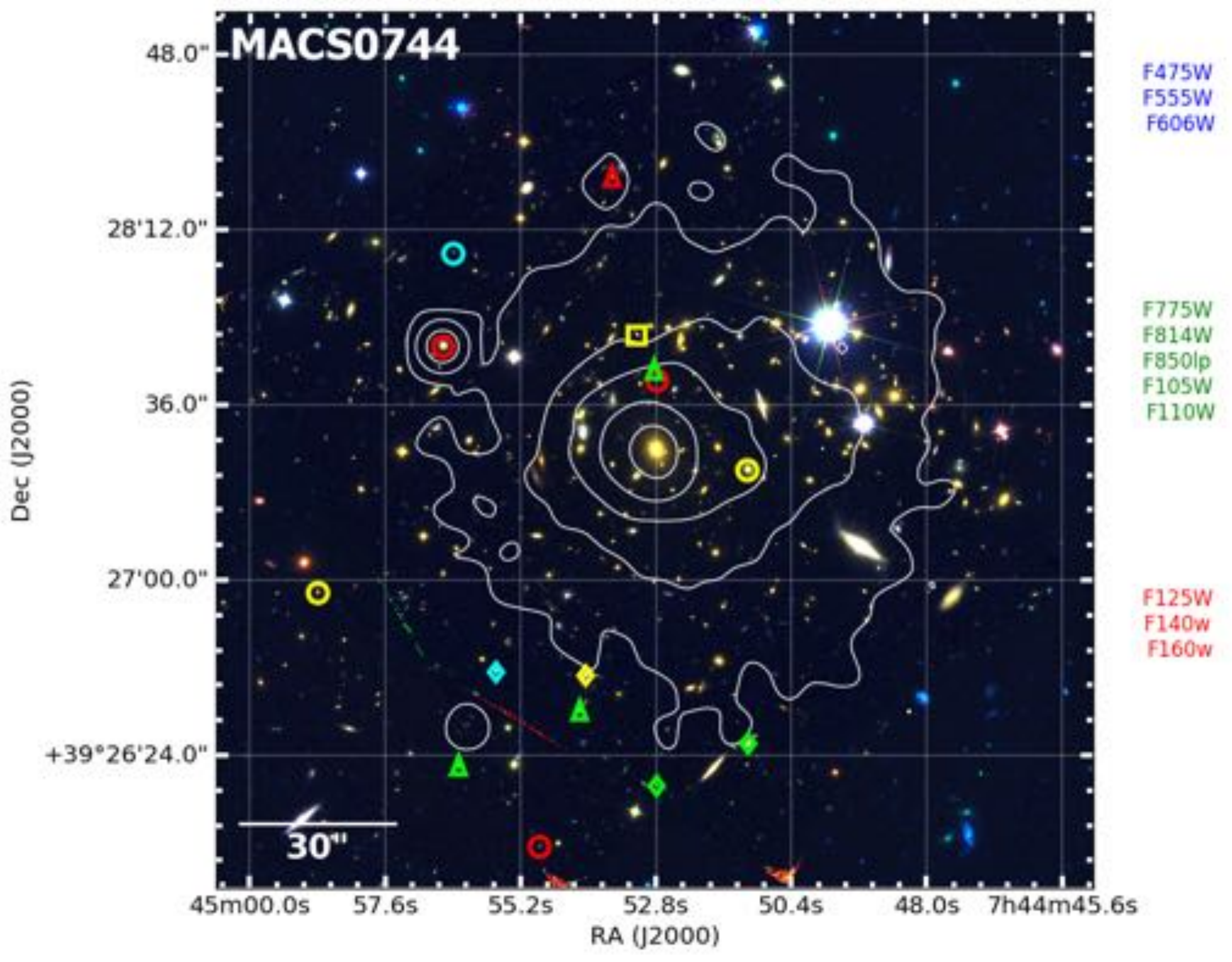}
\caption{}
\end{figure*}

\renewcommand{\thefigure}{\arabic{figure}}

\begin{figure*}[!t]
\centering
\includegraphics[scale=0.35]{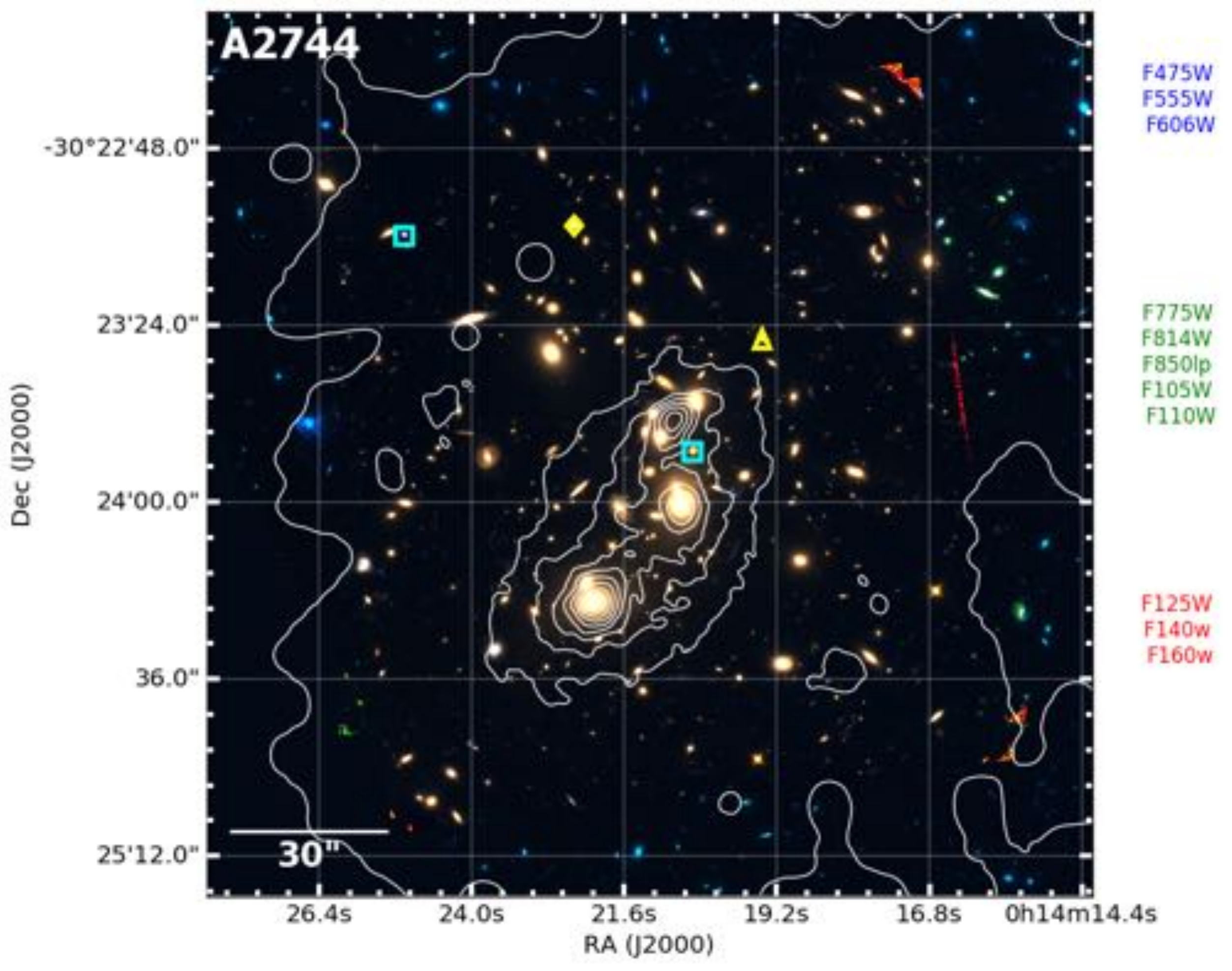}
\includegraphics[scale=0.35]{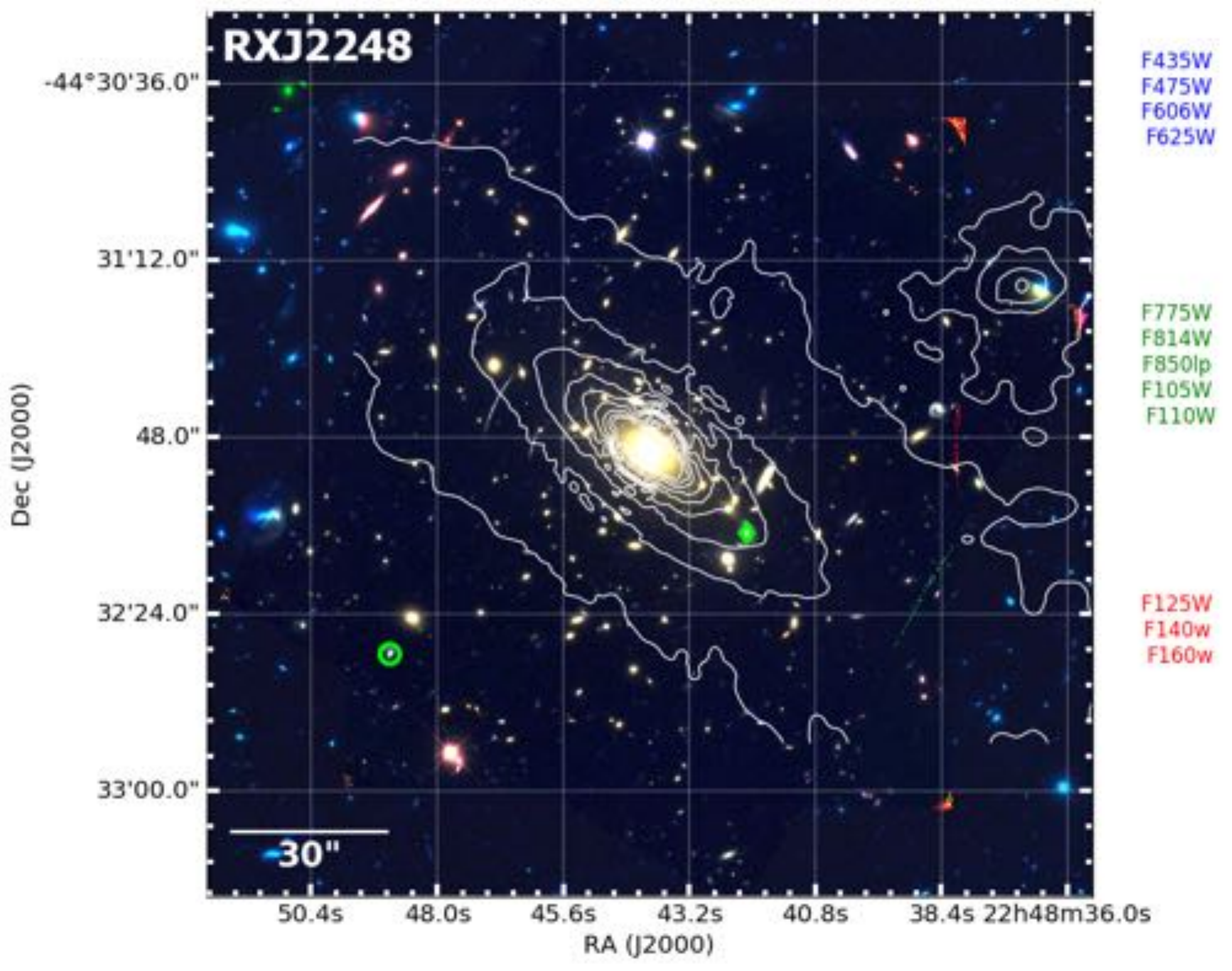}
\includegraphics[scale=0.35]{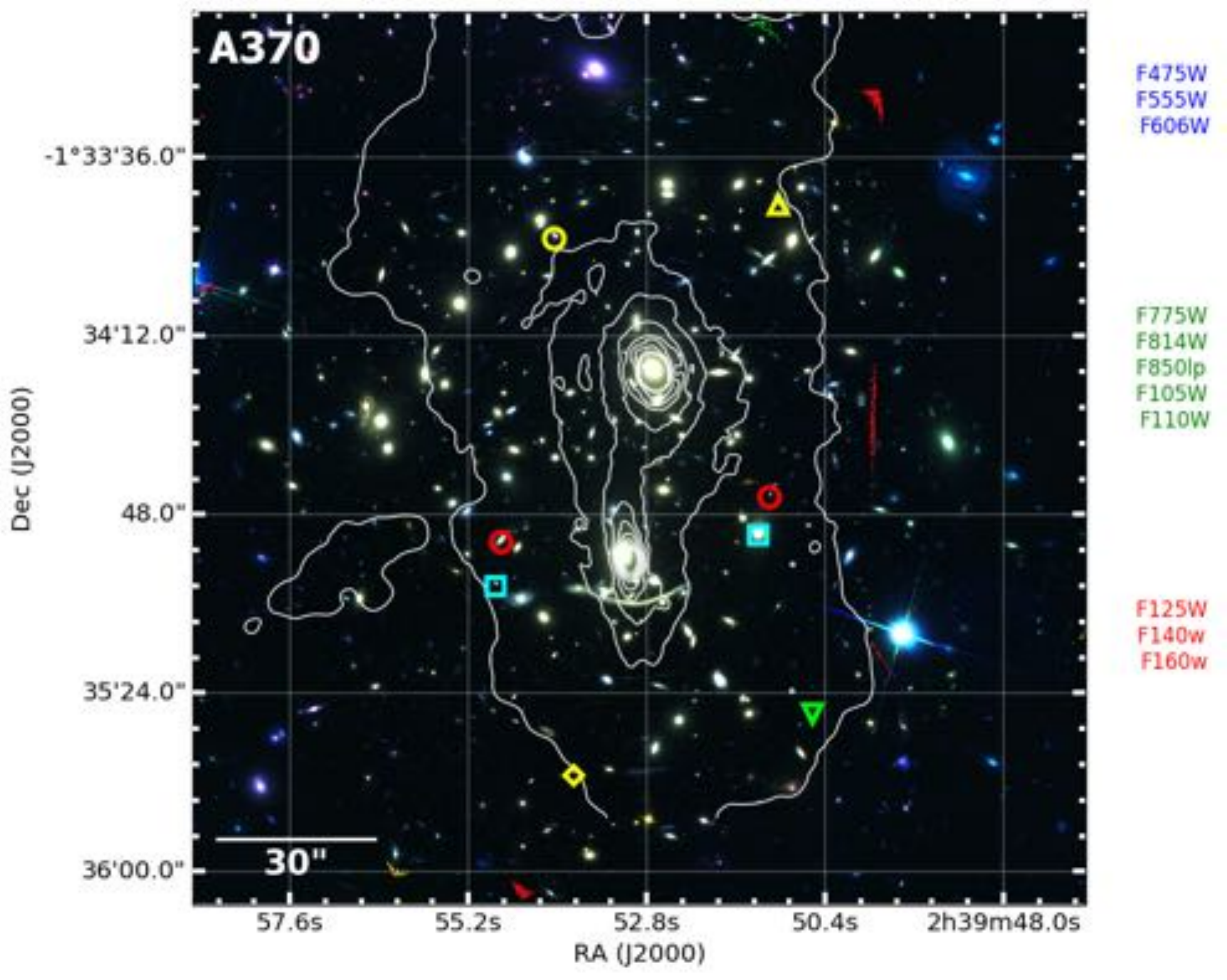}
\includegraphics[scale=0.35]{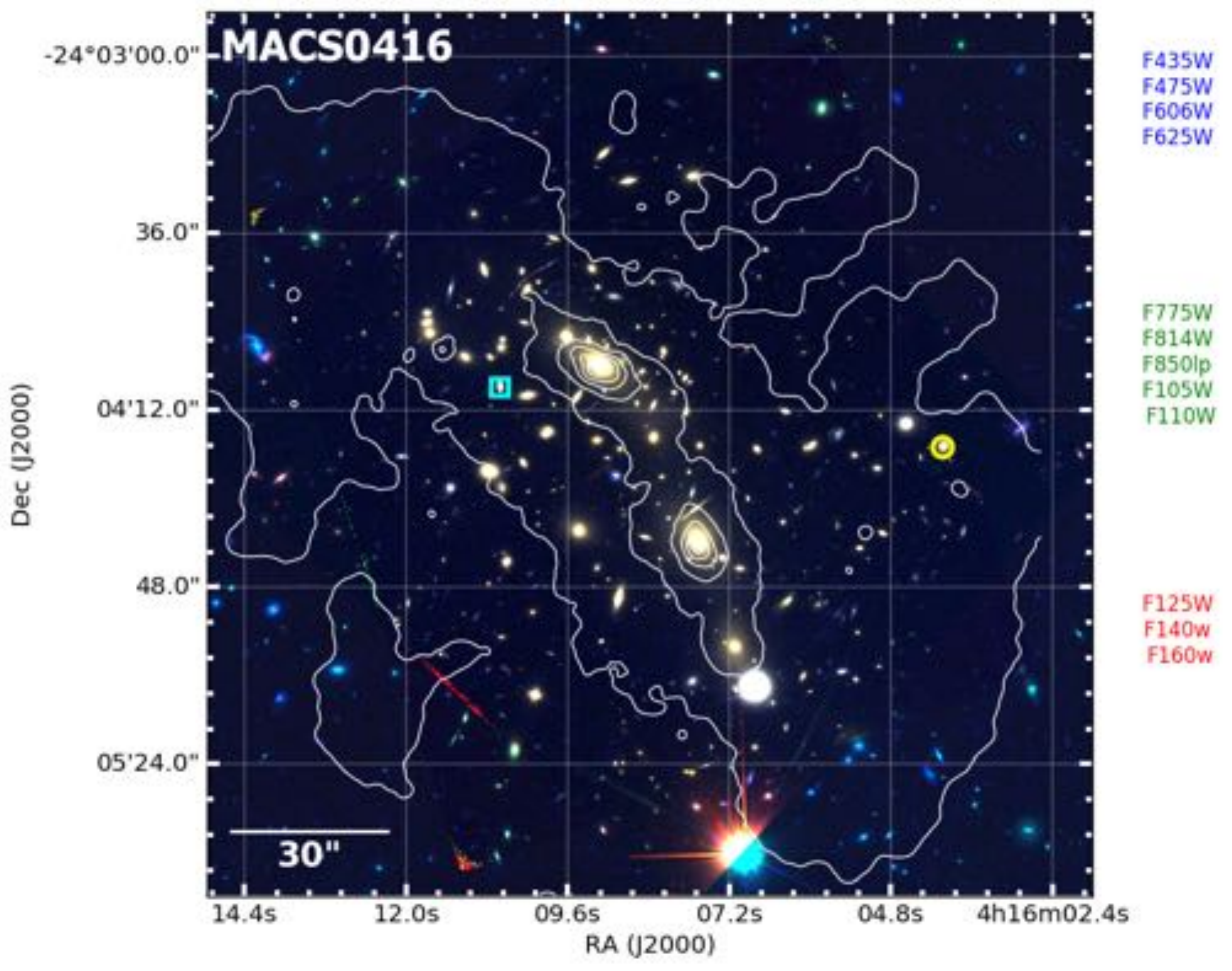}
\includegraphics[scale=0.35]{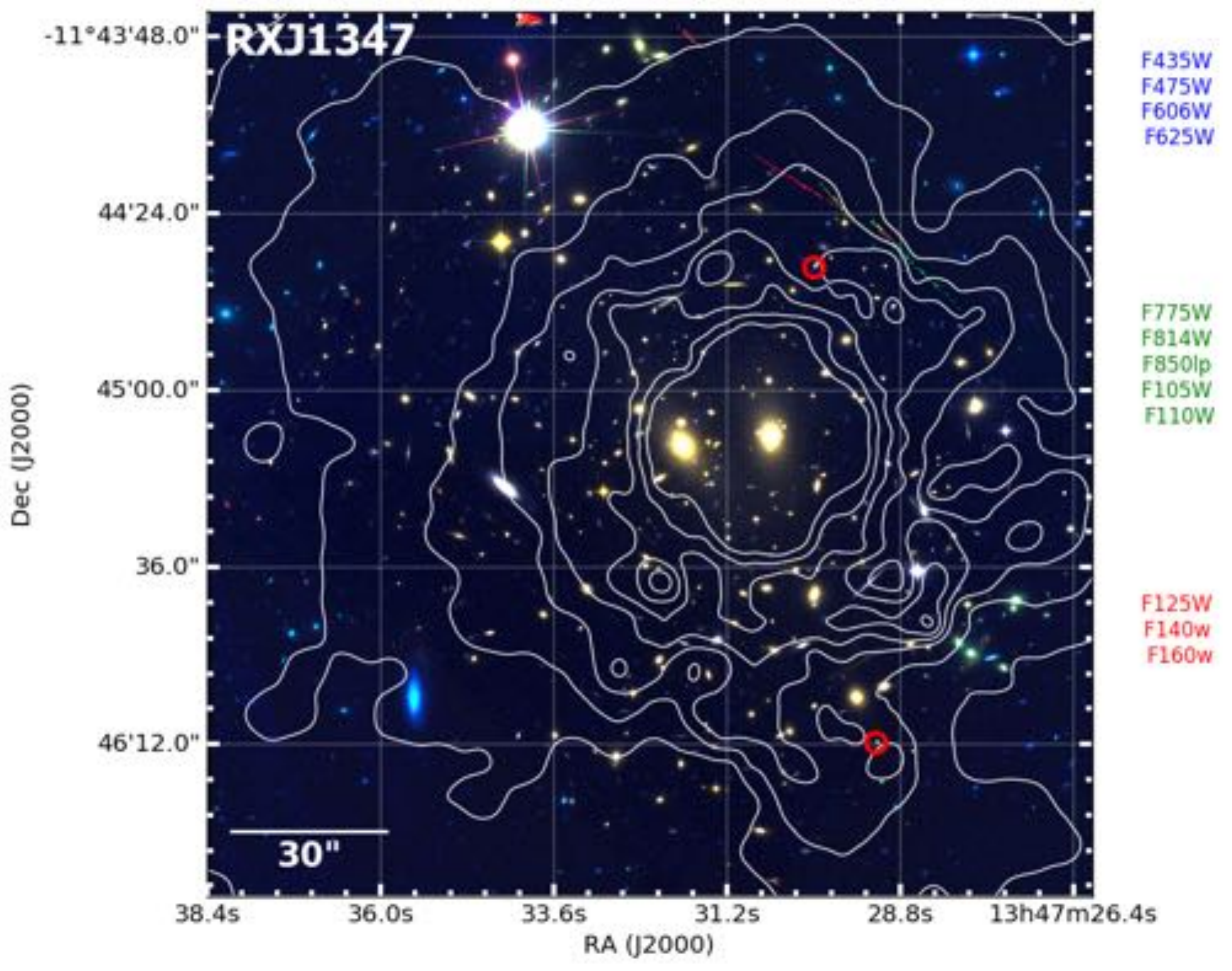}
\includegraphics[scale=0.35]{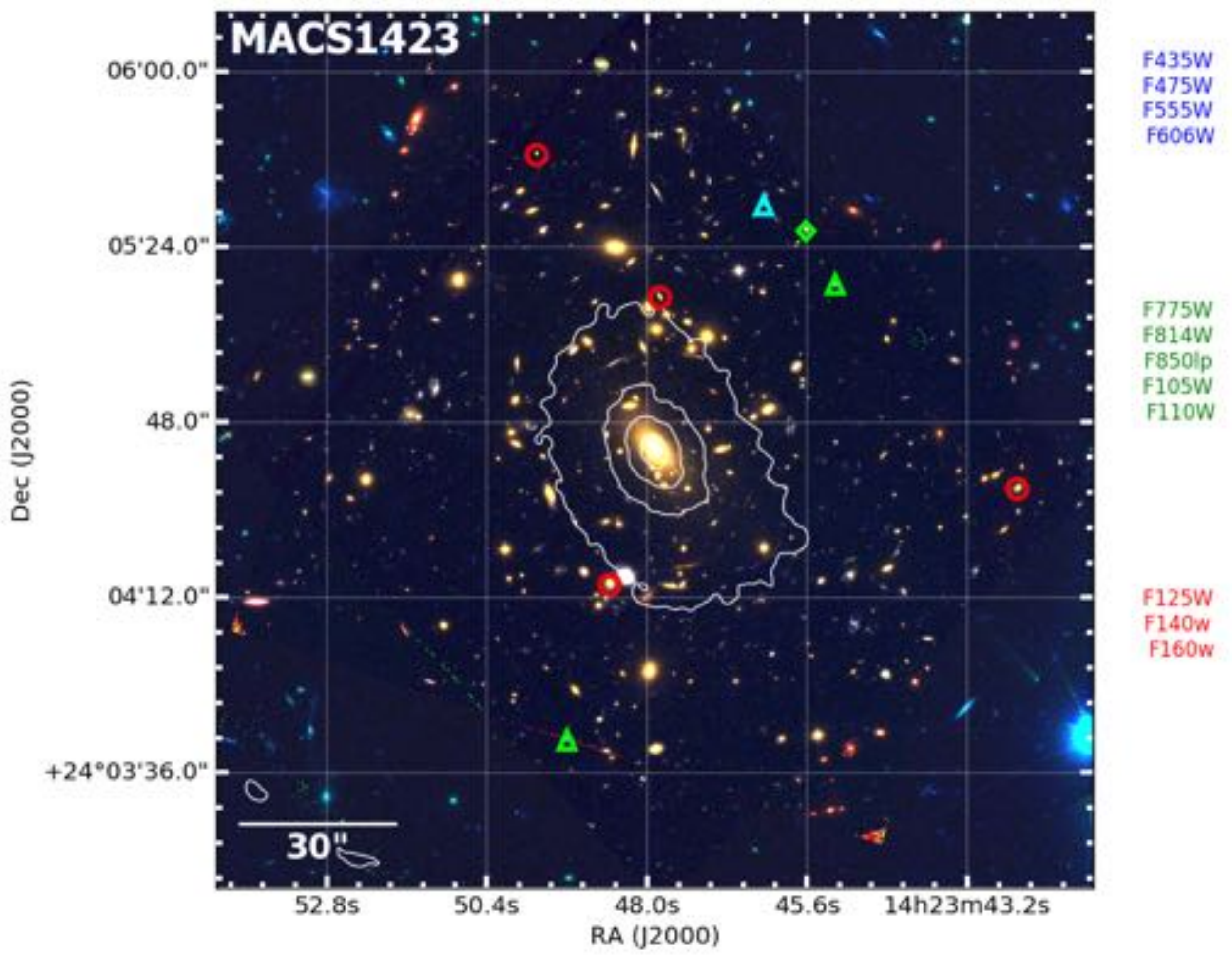}
\caption{Color composite image of  9 GLASS clusters for which surface mass density maps are available. Images are based on the CLASH \citep{postman12} or HFF \citep{lotz16} HST   data. The blue, green, and red channels are composed 
by the filters on the right.  
Surface mass density contours are  overplotted. Contours are  
spaced on a linear scale in the range $10^{-5}-10^{-3}\times10^{12} M_\sun\, kpc^{-2}$.  \Ha emitters  with different \Ha morphologies (different colors)  and experiencing different 
processes (different symbols) are also highlighted. Colors and symbols are as in Figure~\ref{xray}.
\label{surf_mass}}
\end{figure*}

\renewcommand{\thefigure}{\arabic{figure} (Continues)}
\addtocounter{figure}{-1}

\begin{figure*}[!t]
\centering
\includegraphics[scale=0.346]{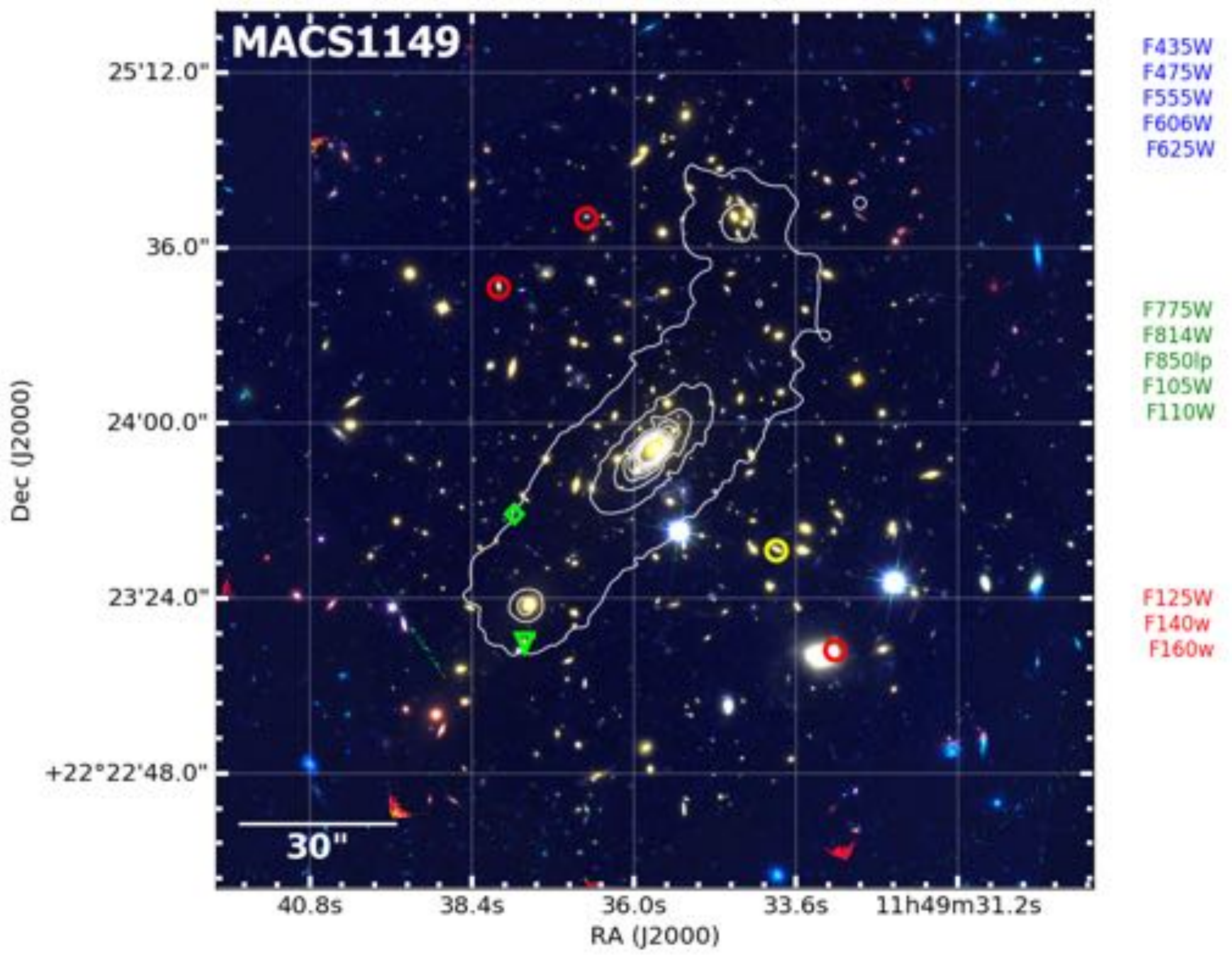}
\includegraphics[scale=0.346]{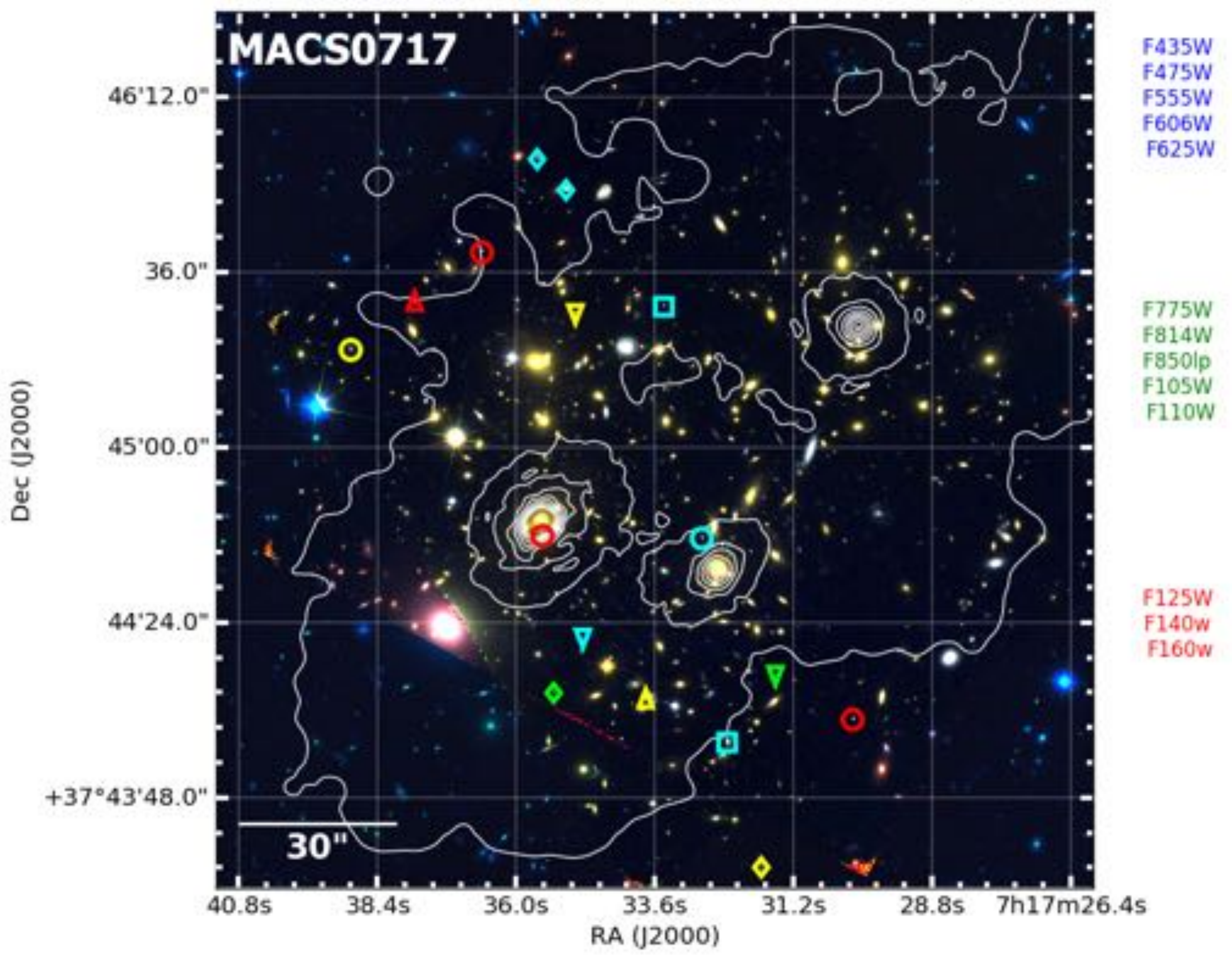}
\includegraphics[scale=0.346]{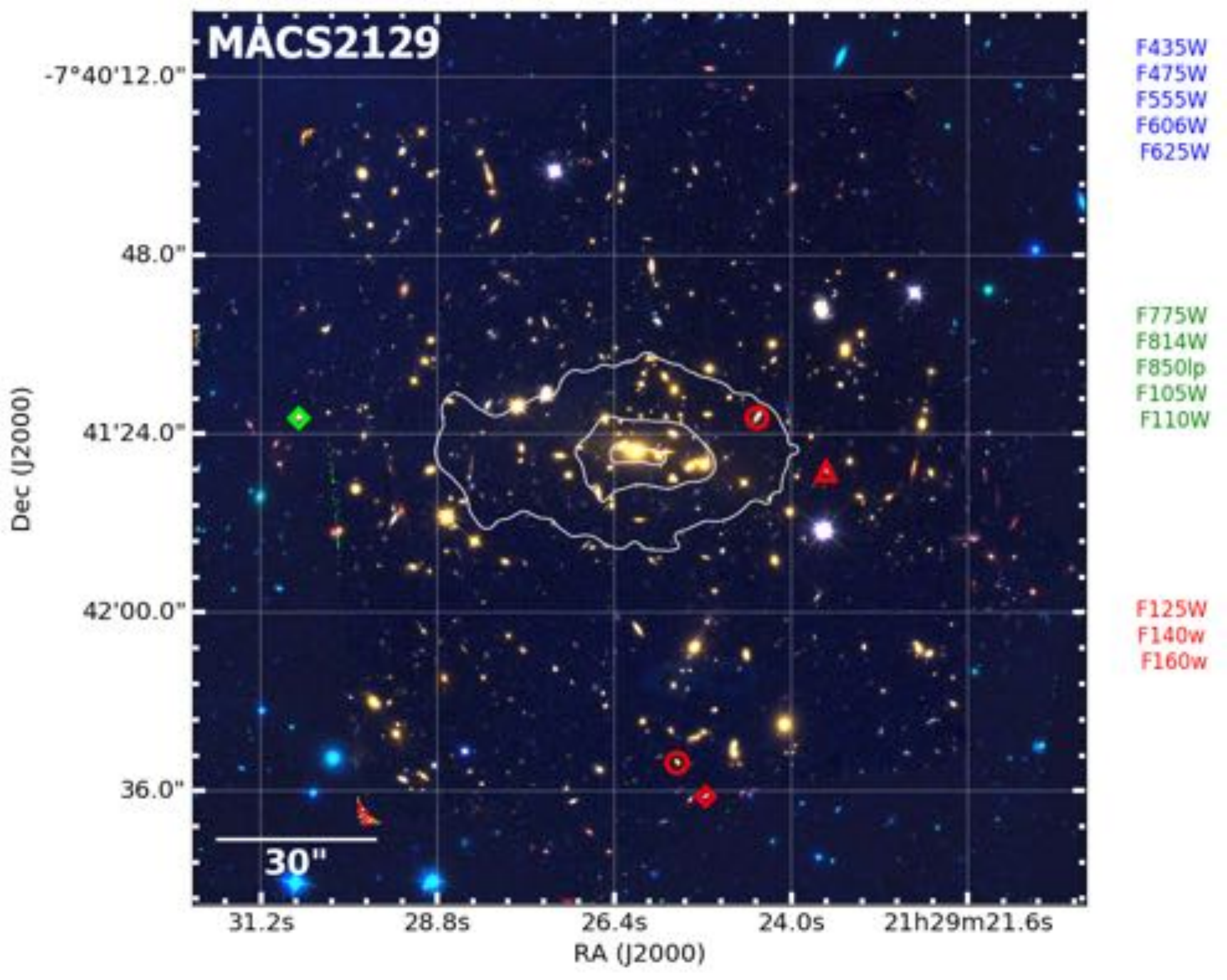}
\caption{}
\end{figure*}

\renewcommand{\thefigure}{\arabic{figure}}


Over the last years, there has been increasing evidence for a correlation between the efficiency of the stripping 
phenomenon and the presence of shocks and strong gradients in the X-ray IGM \citep[e.g.,][]{owers12, vijayaraghavan13}. In  \paperv we found  tentative trends between the X-ray counts and the radial offset,  even 
though  correlations were not supported by statistical tests.

Figure \ref{xray} presents the color composite images of all our clusters along with X-ray maps. Clearly, the clusters in our sample present very different X-ray emission morphologies:  RXJ1347, 
RXJ2248, MACS1423, MACS2129,  and MACS0744 show  quite
symmetric emissions and are relaxed, while A2744, A370, MACS0416, MACS1149,  and MACS0717 have more than one main peak 
and extend along the north south  direction (A370),   the north-west - south-east direction 
(A2744, MACS0717, MACS1149), or the north-east - south-west direction 
(MACS0416). \Ha emitters  with different \Ha morphologies and 
experiencing different processes are  highlighted.

Galaxies with all kinds of \Ha morphologies and also experiencing all the proposed physical 
processes are found in almost all clusters. Due to the low number statistics in each cluster, it is hard 
to detect solid trends with morphology and acting process. In MACS1423, \Ha emitters are 
almost all the  same clustercentric distance, where the hot gas density is nearly constant. 
However, this is not the case for the other relaxed clusters.  In  A2744  and 
MACS0717, characterized by multiple centers, \Ha emitters tend to lie all in the same region of the cluster, and avoid the second peak. 
In MACS0416 there are 
only two \Ha emitters, therefore no solid conclusion can be drawn. The same is true for 
RXJ1347 and RXJ2248. 

It is worth noting that some galaxies are found correspondence of a peak in the X-ray distribution, such as in MACS0744. However, in this case,  the \Ha morphology
seems not to be affected by its peculiar position: indeed it has been visually classified as a galaxy with a regular \Ha morphology where
no strong process is occurring. We remind the reader that the classification has been performed blindly with respect to the cluster properties. 

We note that we cannot know the exact three- dimensional locations of the galaxies with respect to the ICM structures, but in some cases the small projected distances 
from the X-ray peaks and shocks suggest that some galaxies may have recently been overrun by the shock front subcluster gas.
This indicates that a mechanism related to an interaction with these ICM features may be in some cases responsible for either the stripping of the gas or the triggering of the star formation, or both.

Similarly, Figure \ref{surf_mass}  shows the color composite images of 9 of our 
clusters for which the surface mass density maps are available (see \S\ref{sec:clusters}). 
These maps, based on lens modeling, provide an estimate for the total mass density of the cluster, composed  mostly of invisible 
dark matter. Also from these maps the variety of structures in our sample  emerges: 
A370,  MACS0717,   MACS1149, A2744 and MACS0416 
present more than one peak in their distribution, the former extending along the  north-west - 
south-east direction, the latter along the north-east - south-west direction. In contrast,  
 MACS1423, MACS2129,  RXJ2248 and RXJ1347  show nearly 
symmetric mass distribution.  
 
\begin{figure*}
\centering
\includegraphics[scale=0.45]{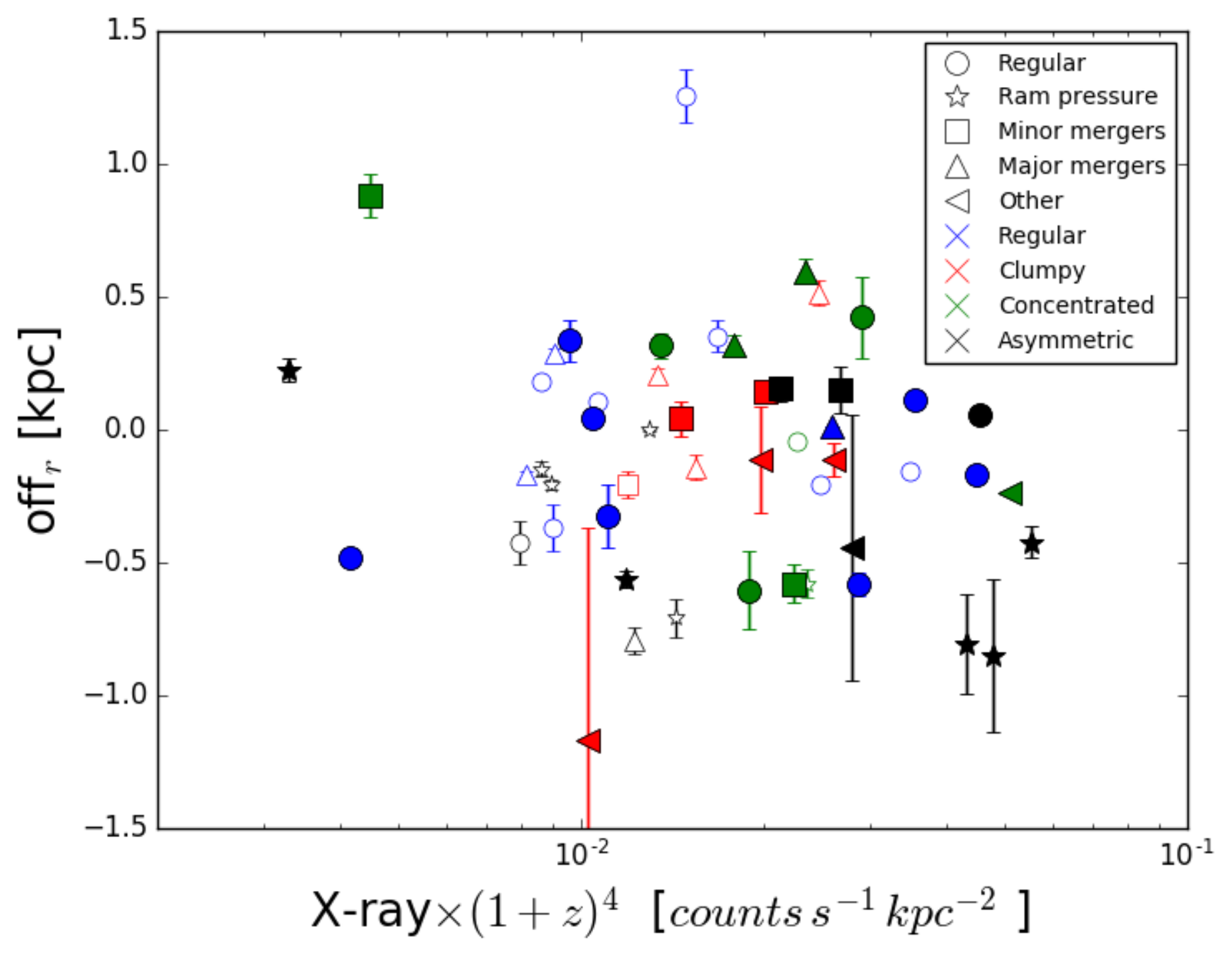}
\includegraphics[scale=0.45]{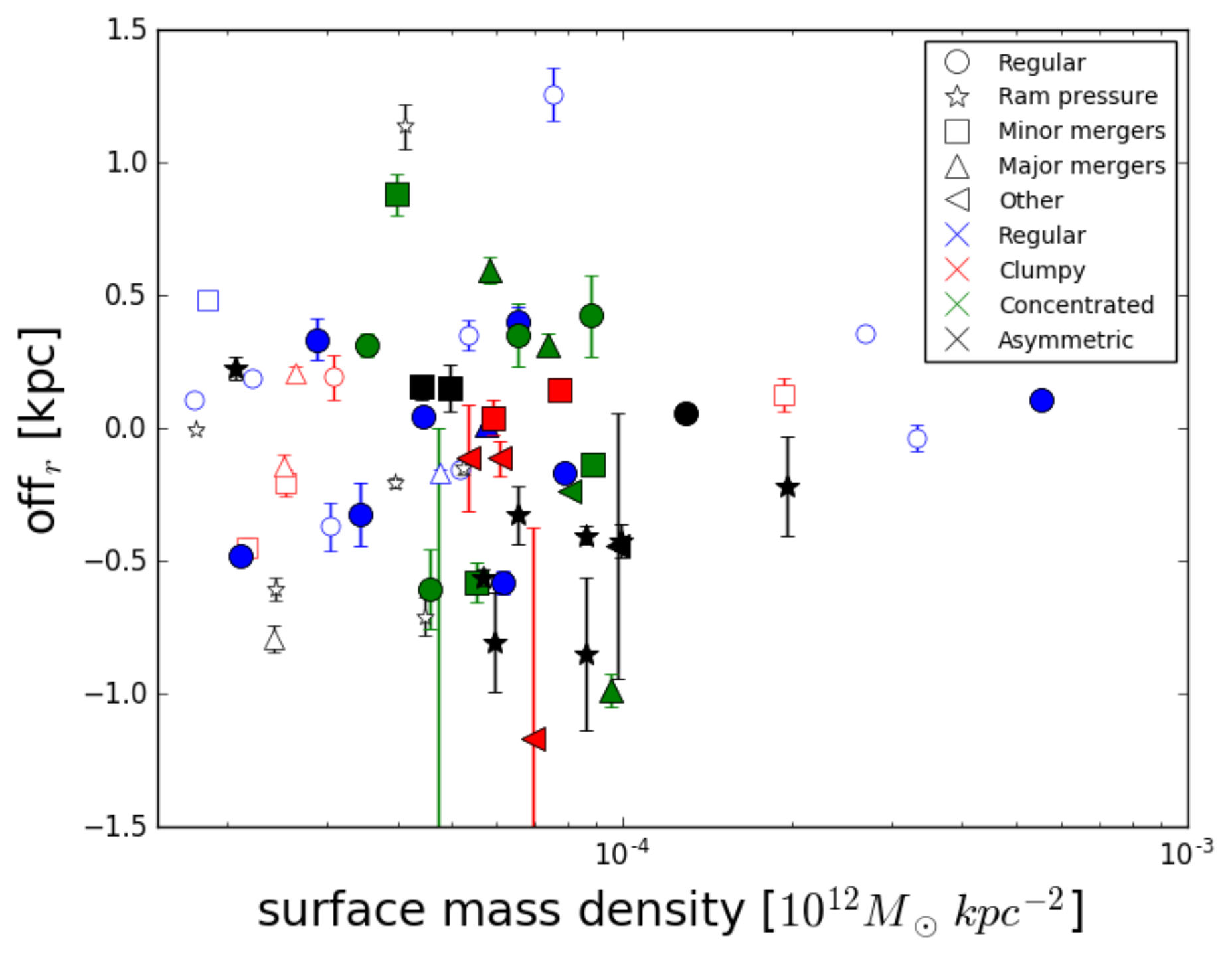}
\caption{Correlation between the radial projected offset and the X-ray emission (left) and surface mass density (right)  for galaxies with different 
\Ha morphology (color)  and 
experiencing different physical processes (symbols), as indicated in the label. Filled symbols represent galaxies in unrelaxed clusters, empty symbols galaxies in relaxed clusters. The radial offset does not correlate with either the X-ray emission or the surface mass density distribution when the whole sample is considered, but it anti-correlates for merging systems. 
\label{offsets_maps}}
\end{figure*}

Figure \ref{offsets_maps} correlates the projected radial offset  with both the  X-ray emission  and the  
surface mass density. X-ray surface brightness has been corrected for cosmological dimming ($\propto (1+z)^4$).
 In both cases, Spearman rank-order correlation tests 
show that no correlation is present between these quantities (Spearman correlation=0.004 
with 80\% significance). However, if we consider only galaxies in unrelaxed clusters, a   weak correlation seems to emerge, 
in the sense that galaxies at higher X-ray counts and surface mass densities tend to have more negative offsets. 
The Spearman correlation test supports these findings. 
This result might suggest that in merging systems X-ray counts are  a proxy for mergers between substructures \citep[see, e.g.][]{poggianti04}, while in the relaxed ones they simply trace the density of the ICM, without inducing 
an alteration in galaxy properties. 

In addition, galaxies in unrelaxed clusters tend to be located systematically at higher X-ray counts and tend to avoid lower surface mass densities than galaxies in all clusters, indicating that in merging systems
the gas temperature and the total dark matter are larger. 


Trends with  \Ha morphology or acting process are hardly detected, both in relaxed and unrelaxed clusters. Few \Ha asymmetric, 
ram-pressure stripped candidates are indeed at high values of the X-ray emission 
\citep[in agreement with ][]{owers12, vijayaraghavan13}, but we find  others  at intermediate values. 
Conversely, not stripped galaxies are found at high values of X-ray emission.

In order to better quantify the impact that the hot gas or the cluster total mass can have on galaxy properties, 
we have investigated the distribution of morphologies and  \Ha morphologies as a function of both X-ray 
emission and surface mass density distribution (plots not shown). 
While at low values of X-ray counts and surface mass density galaxies of all morphological types 
exist, only ellipticals are found at high values of surface mass density, and 
ellipticals and spirals at large X-ray count values.  Focusing on \Ha properties, galaxies with a 
regular \Ha disk seem not to avoid very dense regions, 
where also asymmetric and clumpy objects are found. 

 It would be interesting to investigate the hot gas and surface mass density ranges over which 
the different physical processes take place, but given the small size of our sample, 
significant trends can not be detected.

To conclude, even though some  trends are only tentative, cluster properties like the hot gas density or the 
dark matter distribution seem  to have an impact on the \Ha morphology, and thus on the location of ongoing star formation, only in unrelaxed clusters. 
The lack of strong correlations does not allow us to identify a unique strong environmental 
effect that originates from the cluster center. 

In the next section we will investigate whether some local effects, uncorrelated to the cluster-centric 
radius, play a larger role. 

\subsection{\Ha morphologies as a function of the projected local galaxy density}\label{sec:local}
\begin{figure}
\centering
\includegraphics[scale=0.45]{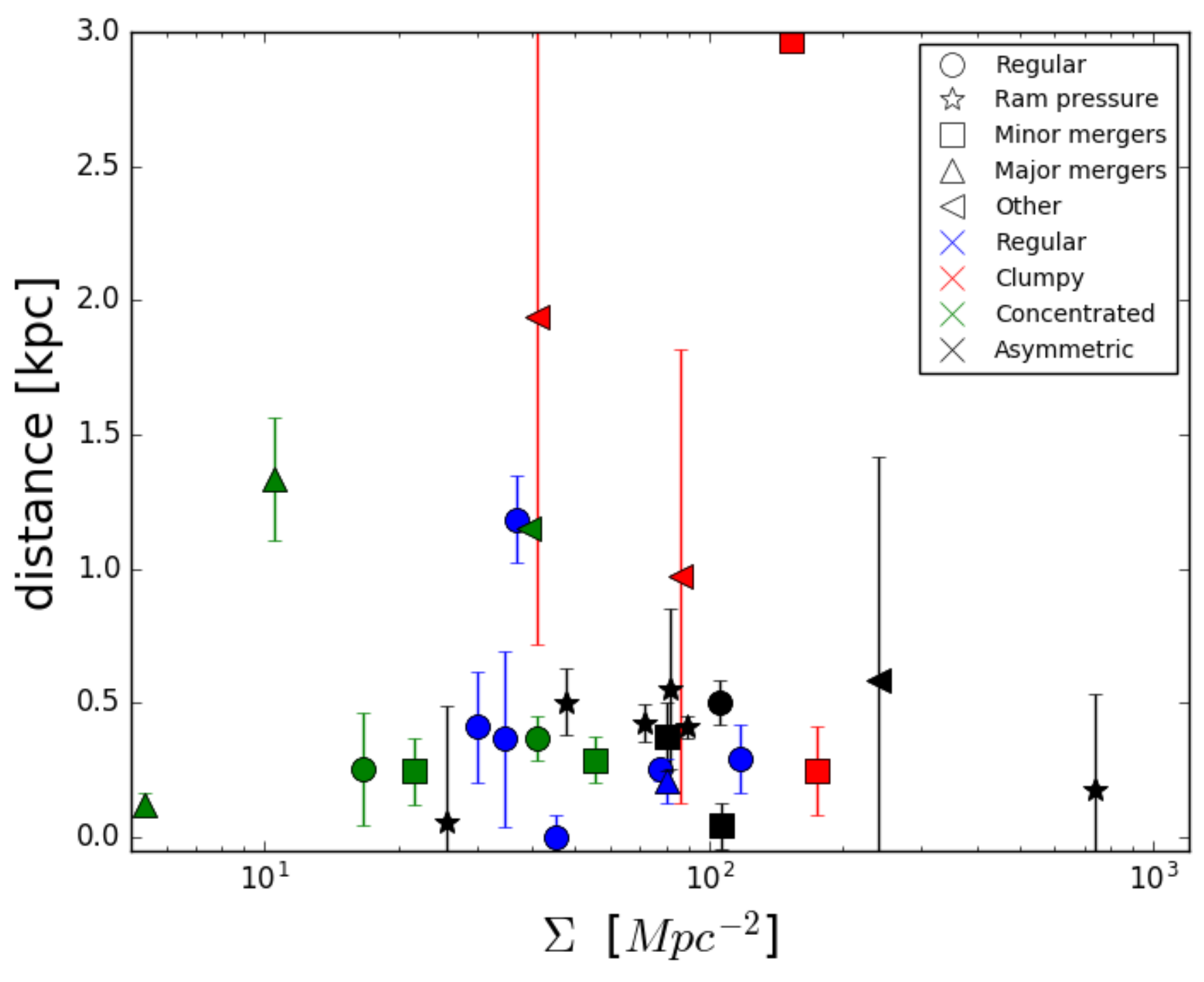}
\caption{Correlation between the 2D distance between the \Ha emission and the continuum emission (F475W filter) and the projected local number density, for galaxies 
 with both PAs. Galaxies with different \Ha morphology 
and experiencing different physical processes are shown using different color and symbols, as indicated in the label. The magnitude of the offset does not seem to correlate with the local density, even though there might be an excess at intermediate values of local density.
\label{LD}}
\end{figure}


Figure \ref{LD} correlates the 2D distance between the peak of the \Ha emission and that of the continuum, 
as traced by the F475W, to the projected local galaxy density. More precisely, it considers 
the absolute value of the offsets (not projected along the clustercentric distance) in the two 
directions (obtained from the two different PAs) and, for the galaxies with both PAs, the 
real distance between the two peaks, obtained by combining the offsets.

The magnitude of the offset does not seem to correlate with the local density (Spearman 
correlation tests are always inconclusive), even though there might be an excess at 
intermediate values of local density. Nonetheless, some segregation effects between local density and \Ha 
properties (different colors in Figure~\ref{LD}) are visible. 

Galaxies with concentrated \Ha\  seem to be preferentially found at lower densities, while galaxies with 
asymmetric \Ha\  seem to prefer denser environments. Galaxies with regular and clumpy \Ha are found at intermediate values 
of local density. 
K-S tests confirm that each population is drawn from a different parent distribution with high significance levels (>90\%), 
except for galaxies with regular and clumpy \Ha. 

Regular processes seem to operate at low to intermediate densities, as is also true for mergers. 
In contrast, ram-pressure stripping and  unidentified 
processes tend to operate also at higher densities.   A K-S test can reject the null hypothesis
 that a regular process and ram-pressure stripping are drawn from the same distribution at $\sim90\%$ confidence.


To conclude, despite the statistics limited to 4 out of 10 GLASS clusters, trends with local densities are stronger than trends with the other tracers.

\subsection{sSFRs as a function of environment}\label{sec:ssfr}
Understanding the origin of the trends of star formation with cluster properties  represent a significant 
step forward toward comprehending the link between galaxy evolution and 
environment. If galaxy properties depend on the mass of the system where they reside or have resided 
during their evolution, there should be a connection between the trends 
observed and the way cosmological structures have grown in mass with redshift.

In \cite{vulcani10} and \paperxi we have found that the SFR-mass relation 
depends on environment: while many galaxies in clusters can be as star-forming as 
galaxies in the field, in the more massive systems a population of galaxies with a reduced 
SFR at fixed mass is detected. This result indicates that  some cluster-specific 
processes that suppress star formation are taking place.  

\begin{figure*}
\centering
\includegraphics[scale=0.45]{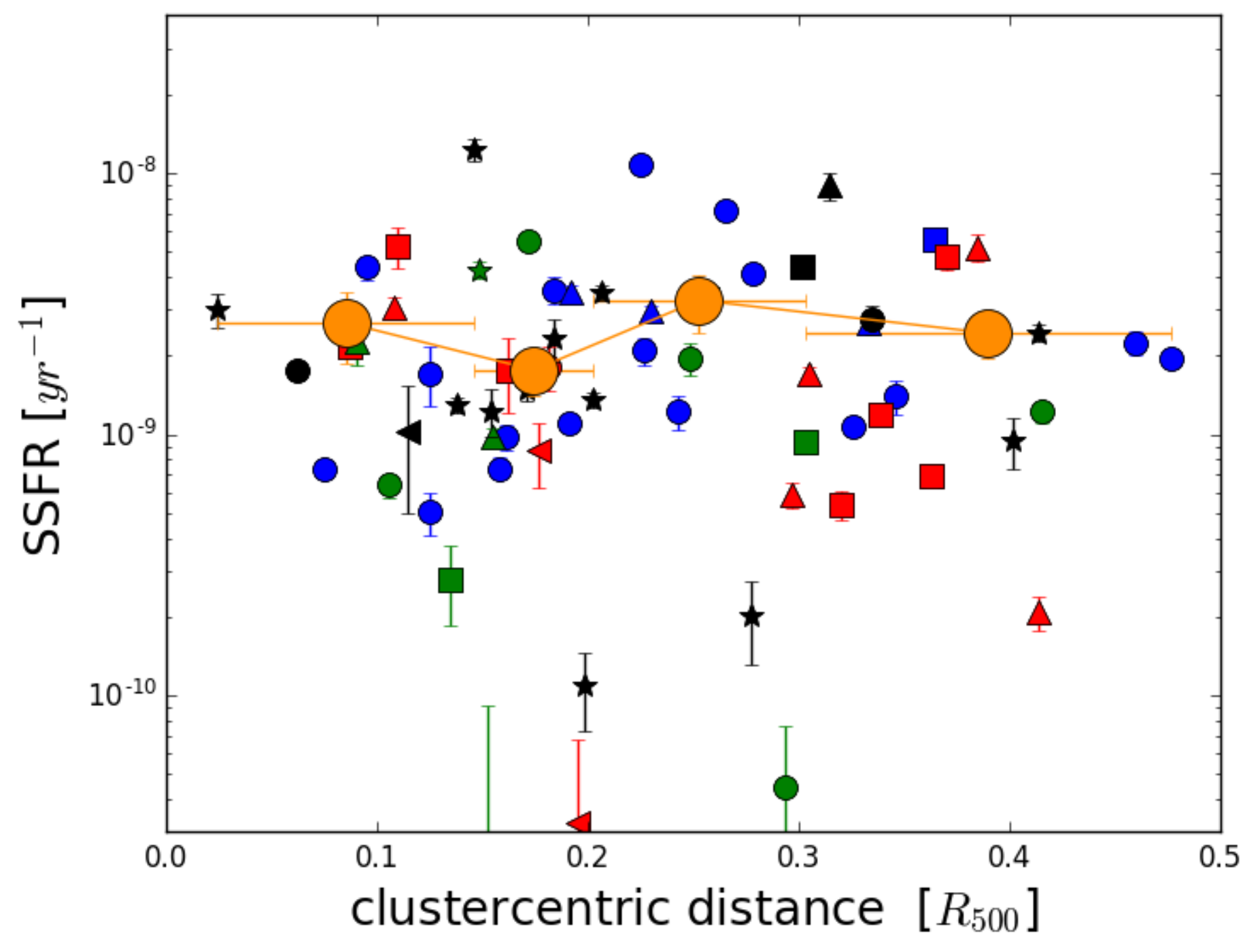}
\includegraphics[scale=0.45]{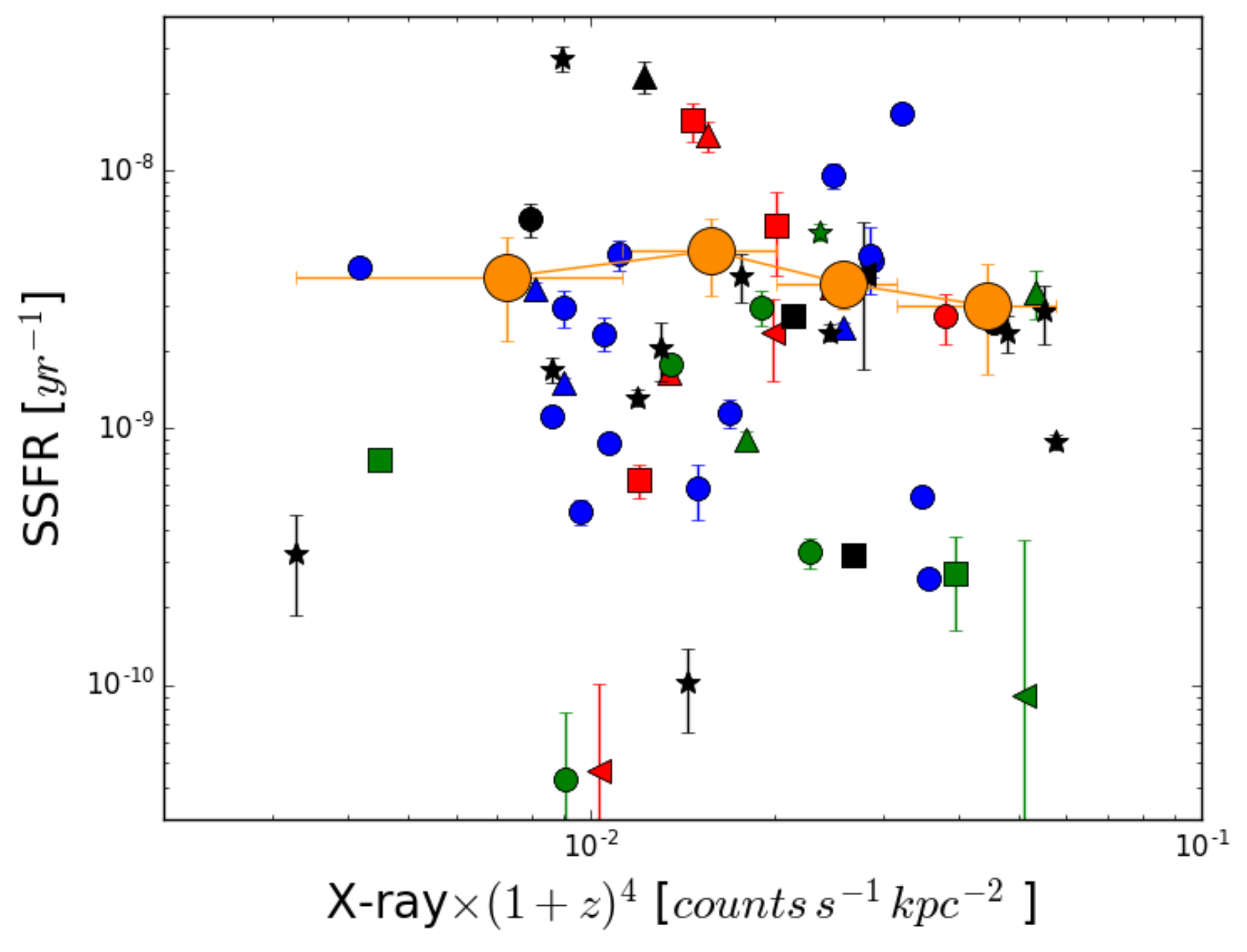}
\includegraphics[scale=0.45]{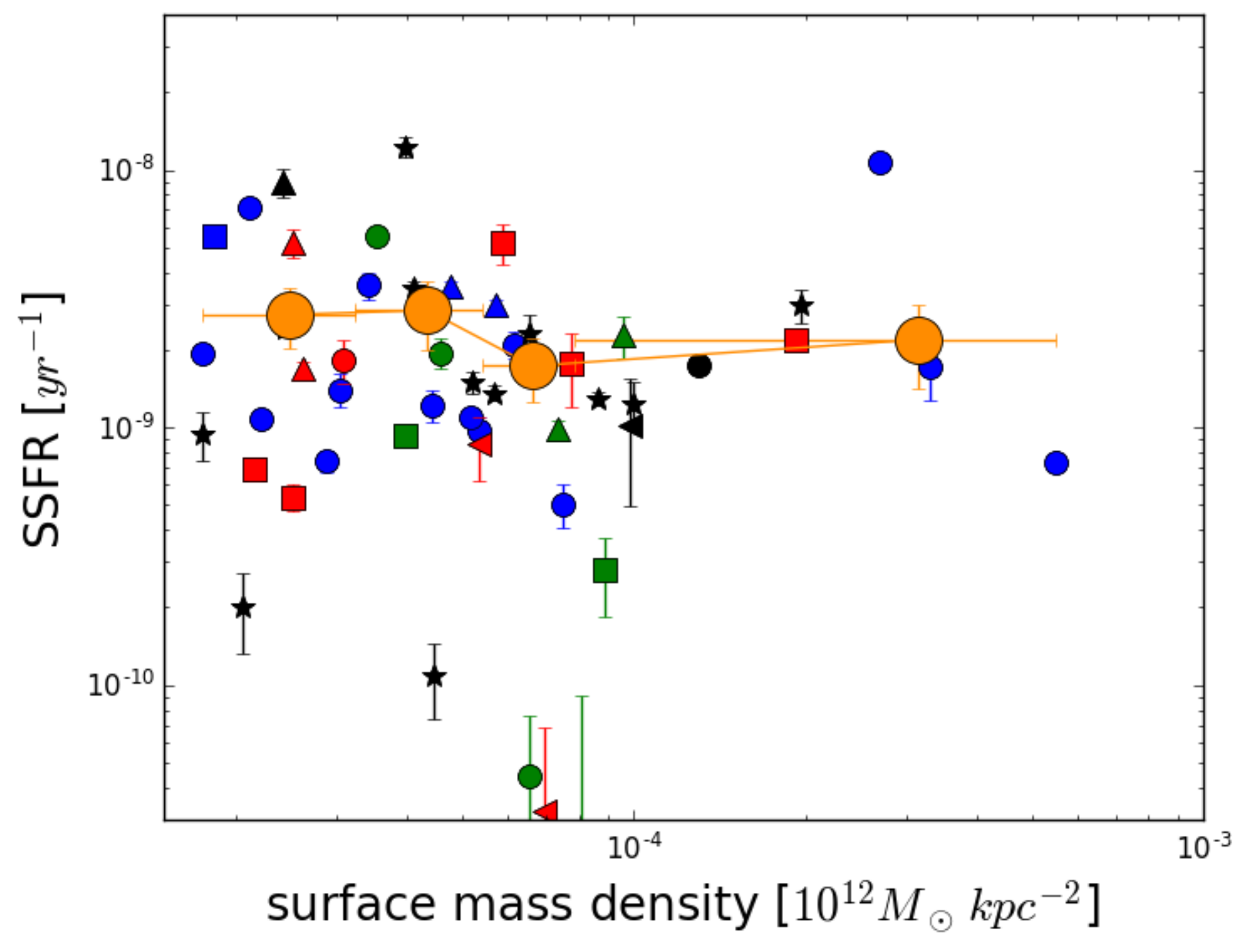}
\includegraphics[scale=0.45]{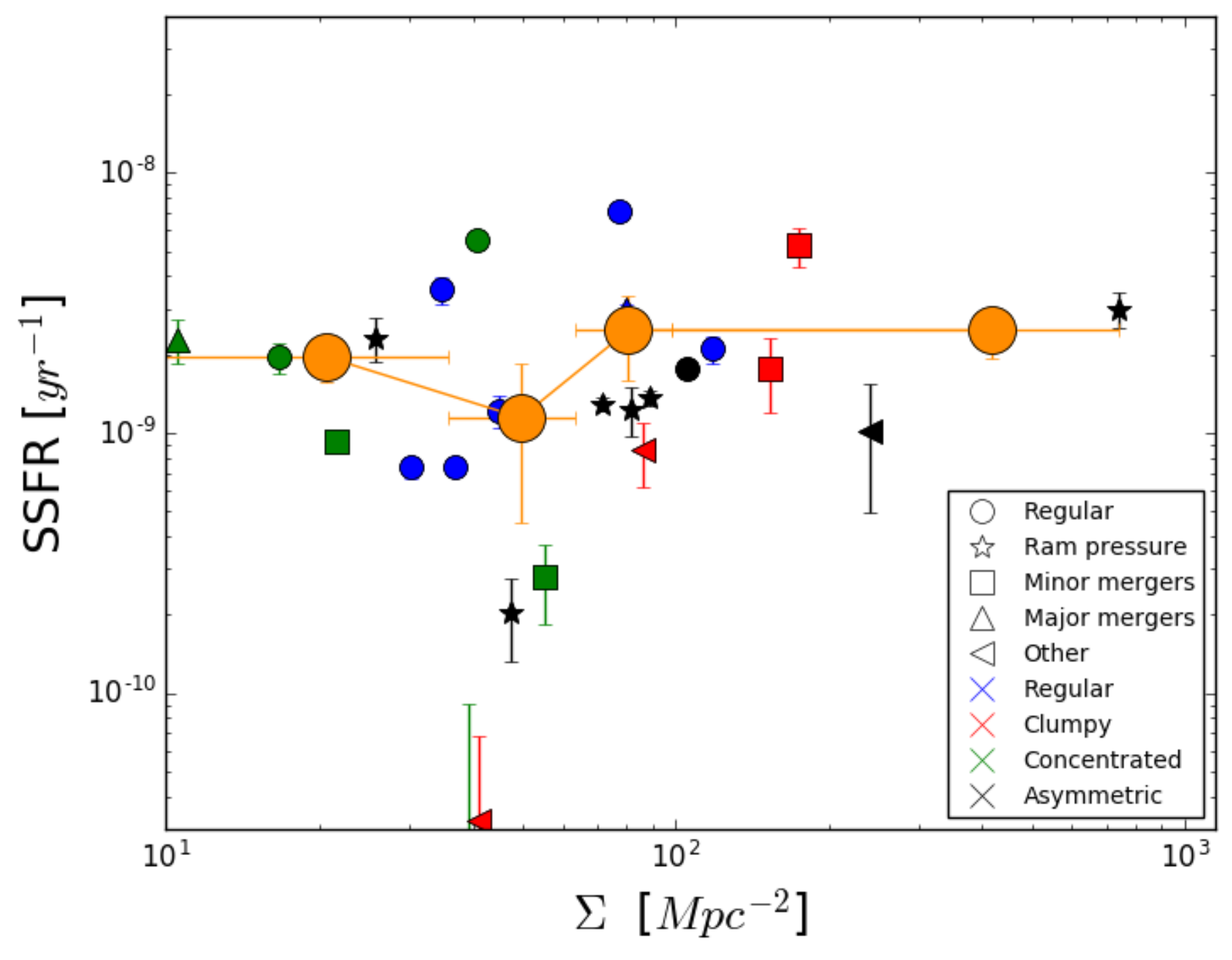}
\caption{sSFR  as a function of different parameterizations of environment for  galaxies with different 
\Ha morphology and 
experiencing different physical processes, as indicated in the label. Errors on the individual measurements are typically smaller than the symbols. 
Big orange dots with errorbars  indicate 
 mean values in four equally-populated bins of the considered quantity. 
Upper left panel: clustercentric distance (all clusters), upper right panel: X-ray counts (all clusters), bottom left panel: 
 surface mass density values (nine clusters), bottom right panel: projected local density (four clusters). All trends appear flat,
 indicating that the average sSFR in star forming galaxies does
not depend on either global or local environment.   
\label{sSFR_glob}}
\end{figure*}

Here we just focus on clusters, 
and search for differences in the typical star forming properties for galaxies living in 
different conditions. 
To remove the influence of the stellar mass, we consider the Specific Star Formation Rate 
(sSFR), defined as the SFR per unit of galaxy stellar mass. 
As shown in Figure~\ref{sSFR_glob} the mean sSFR seems  to depend on neither global nor local environment. 
 
 Considering separately galaxies with different \Ha morphology and/or experiencing different processes, 
no trends emerge. 

Putting together these results and those presented in \cite{vulcani10} and \paperxi, we conclude
that while differences emerge when comparing the most different environments in the Universe (clusters vs. field),
the  properties of the clusters do not seem to  strongly influence the star formation in the cluster members.  However,
all these trends will need to be confirmed  or refuted  with larger number statistics.

We stress that we do not have all measurements for all our clusters, so our sample could be affected 
by incompleteness.  Indeed, in principle, results from the clusters not included in our sample could 
differ from those presented here. Nonetheless, our findings are based on a random subset of the whole 
sample, therefore, for each measurement, the incompleteness 
is not related to cluster properties, so we do not expect strong biases. 

We also note that our lack of trends is most likely due to the fact that,  as discussed in \paperxi, our 
sample  has been assembled by selecting visually detected \Ha emitters and therefore  includes only highly star forming objects which most likely are at the first infall. 
Their SFR might therefore not been yet affected by the dense cluster environment. To properly
characterize the effect of the environment one should focus on galaxies that have been part of the system 
for a long time  and reaching lower  SFR levels.

\subsubsection{Comparison with previous work}

Many authors have investigated the relation between star formation in  galaxies as 
a function of environment, mainly focusing on the star forming fractions 
\citep[e.g.,][]{poggianti06, dressler13, zabludoff98, biviano97, smail98}.
For our sample it is not straightforward to perform a similar analysis, 
since the GLASS dataset does not yield redshifts for passive galaxies.
Nonetheless, the fact that we do not find trends with the environment is overall in agreement with previous analyses. 

Overall, at intermediate redshifts ($z\sim1$), there is still no consistencies between different works:  while some studies show a lower SFR or sSFR in denser regions compared to less-dense ones
\citep[e.g.,][]{patel09, muzzin12}, some provide evidence for  flat relations \citep[e.g.,][]{grutzbauch11, scoville13}, and there are even reports of a correlation between  star formation activity and density \citep[e.g.,][]{elbaz07, cooper08, welikala16}.

\cite{finn05}, investigating 3 clusters at $z\sim0.75$, found that  the fraction of star-forming 
galaxies increases with projected distance from the cluster center and  decreases with 
increasing local galaxy surface density, but the average SFR does not \citep[see also][]{dressler80, dressler16}. Comparing galaxies in 
clusters at z$\sim0$ and $z\sim0.5$, \cite{poggianti08} found that in both nearby 
and distant clusters, higher density regions contain proportionally fewer star-forming galaxies, and the 
average [OII] equivalent width of star-forming galaxies is independent of 
local density. Their results suggest that at high z the current star formation activity in star-forming 
galaxies does not depend strongly on global or local environment.

Similarly, at $z<0.1$, \cite{balogh04} found that the relative numbers of star-forming and quiescent 
galaxies varies strongly and continuously with local density \citep[see also][]{kauffmann04, baldry06, darvish16}. However, amongst 
the star-forming population, the distribution of the equivalent width of \Ha, which can be used as a
proxy for the strength of the specific star formation,  is independent of environment 
\citep[see also][]{tanaka04, wijesinghe12}. In contrast, \cite{vonderlinden10} found a marked 
anticorrelation between star formation and radius, which  
is most pronounced for low-mass galaxies and is very weak or absent beyond the virial radius. 
Discrepancies among the different studies can be explained in terms of the 
different SFR completeness limits reached: in \cite{vonderlinden10}  the decline in SFR is driven largely 
by the inclusion of green galaxies into the sample of star-forming galaxies 
with low levels of star formation, which are instead missing in \cite{balogh04} and \cite{tanaka04} samples.

\section{Summary and Conclusions}\label{sec:conc}

Building on our previous work described in \paperv and  \paperxi, we
have carried out a detailed investigation of the spatial distribution
of star formation in galaxies at $0.3<z<0.7$, as traced by the \Ha
emission in the 10 GLASS clusters.

\Ha maps were produced taking advantage of the WFC3-G102 data at two
orthogonal PAs.  We have visually selected galaxies with \Ha in
emission and, based on their redshift, assigned their membership to
the cluster.  Following \paperxi, we have computed SFRs, and visually
classified galaxies, paying particular attention to their broad-band
morphology, and their \Ha morphology. The new scheme introduced in
\paperxi visually categorizes galaxies according to the main process
that is affecting the mode of star formation. Ours is clearly a
qualitative and approximate classification scheme, considering that
multiple processes might be simultaneously at work, but we believe
there is merit in categorizing in a self consistent manner the
diversity of morphological features across environments.

In this paper we have correlated the \Ha morphology with environmental conditions
in which galaxies are embedded, focusing on the clustercentric
distance, the hot gas density from X-ray emission, the total surface mass density from gravitational lensing
 and the local
projected number density, to give a better insight on the role of the cluster
environment in driving galaxy transformations.

Our main results can be summarized as follows. 

\begin{itemize} 

\item \Ha emitters can be found both close to the cluster center and
  up to 0.5$r_{500}$, which is the maximum radius covered by
  GLASS. The radial projected offset between the peak of the \Ha
  emission and the peak  in the continuum (as traced by the F475W
  filter) is negative for 60\% of \Ha emitters, indicating that for
  most of them ionized gas is preferentially displaced away from the
  cluster center. In contrast, the distribution of the tangential
  offset shows no preferential direction. This result is solid at $\sim 2\sigma$ level. 

\item In order to quantitatively test the hypothesis that ram-pressure
  stripping is the main driver of the observed radial offsets between
  \Ha and the continuum, we compare with the \citet{Jia15} numerical
  simulations. As expected, given the small clustercentric radius of
  observation of the \Ha emitters, they consist of the 25\% most
  radial orbits found in cosmological simulations. Assuming that the
  direction of the offset between \Ha and continuum can be taken as
  proxy for the direction of motion at the time of observations, we
  find that the observed distribution of directions is consistent with
  the expectations for infalling galaxies in cosmological simulations.
  The agreement improves when we consider  only galaxies visually
  classified as undergoing ram-pressure stripping, providing
  quantitative support for our interpretation of the morphology.

\item Our clusters cover a wide range of morphologies: some of them
  are relaxed, while others are merging systems presenting very
  asymmetric configurations, as probed by the different surface mass
  density distributions and X-ray emissions.  Galaxies characterized
  by all kinds of \Ha morphologies and experiencing the different
  processes can be found in almost all clusters and significant trends cannot 
  be detected using this relatively small data sample. However, when considering
  only unrelaxed clusters we find that galaxies 
  found in correspondence of a peak in the X-ray and surface mass distributions
  have more negative offsets.
  This indicates that a mechanism related to an interaction with these 
  ICM features may be in some cases responsible for 
  an alteration in the star forming properties.

\item Whereas the amplitude of the offset between the peak of the \Ha
  emission and the peak in the continuum (as traced by the F475W
  filter) does not depend on local density,  we recovered some hints that \Ha
  morphologies do. 
  Galaxies with concentrated \Ha\  seems to be preferentially 
  found at lower densities, while galaxies with 			
  asymmetric \Ha\  might prefer denser environments. Galaxies with regular and clumpy 
  \Ha are found at intermediate values of local density. 
  K-S tests support these findings. 
  In addition, as expected, mergers are found at low to intermediate
  densities. In contrast, ram-pressure stripping and unclassified
  processes tend to operate at higher densities.

\item The most statistically significant result is that galaxies with
  asymmetric \Ha distribution, interpreted as signatures of recent ram
  pressure stripping, are preferentially found within 0.3 $r_{500}$,
  at higher local density conditions and higher X-ray counts and have
  a negative radial projected offset, i.e. the peak of the \Ha
  emission is pointing away from the cluster center with respect to
  the continuum emission.

\item Overall, the average sSFR in star forming galaxies does 
  depend  on neither global nor local  environment.  These findings will have to be confirmed or refuted by larger number statistic, but, if true, they  suggest that
  the  properties of the clusters are not able to strongly affect the star formation in  clusters.
  However, it is important to stress that our 
sample includes only highly star forming objects which most likely are at the first infall. 
Their SFR might therefore not been yet affected by the dense cluster environment. To properly
characterize the effect of the environment one should focus on galaxies that have been part of the system 
for a long time. 
Galaxies with concentrated \Ha are preferentially found at lower densities, while galaxies with 
asymmetric \Ha prefer denser environments. Galaxies with regular and clumpy \Ha are found at intermediate values 
of local density. 
K-S tests confirm that each population is drawn from a different parent distribution with high significance levels (>90\%), 
except for galaxies with regular and clumpy \Ha.

  \end{itemize}

Based on these observations, we conclude that in clusters the
population of star-forming galaxies, as traced by the \Ha emission, is
very heterogenous. Although living in the most extreme environments of
the Universe, a considerable fraction of galaxies still are not
affected by the surrounding conditions and present regular \Ha
morphologies not affected by any strong physical process. Nonetheless,
many galaxies respond to the extreme conditions in which they are
embedded, especially those in unrelaxed clusters. They show 
torqued, asymmetric, clumpy \Ha morphologies. Many
different processes are thought to be responsible for these
observations and no unique physical process emerges as
dominant. Overall, the most evident trends have been detected with the
local density, suggesting that local effects play a larger role than
those correlated to the clustercentric radius. Such effects weaken
potential radial trends.

Following \citet{dressler80}, in the past several years significant
evidence has been accumulated that several of the main
galaxy properties, such as the galaxy mass, the red galaxy population,
and the morphological types of galaxies, are better correlated with
the local environment  than the global environment.  Both at $z\sim0.6$ and
$z\sim0$, \cite{vulcani12} found that the shape of the galaxy stellar
mass function depends on local density, while variations with the
global environment (intended as cluster vs. field) are very subtle
\citep{vulcani13, calvi13}.  In local clusters, none of the
characteristics of the colour-magnitude red sequence (slope, scatter,
luminous-to-faint ratio, blue fraction and morphological mix on the
red sequence) depends on global cluster properties connected with
cluster mass, such as cluster velocity dispersion and X-ray
luminosity. In contrast, all of these characteristics vary
systematically with the local galaxy density \citep{valentinuzzi11}.
In addition, the fractions of spiral, S0 and elliptical galaxies do
not vary systematically with cluster velocity dispersion and X-ray
luminosity \citep{poggianti09}, while a strong morphology-density
relation is present \citep{fasano15}. Moreover, \cite{balogh04} found
that the red fraction of galaxies is a strong function of local
density, increasing from $\sim10-30\%$ of the population in the lowest
density environments to $\sim70\%$ at the highest densities, while
within the virialized regions of clusters it shows no significant
dependence on cluster velocity dispersion.  Also, \cite{martinez08}
found that bright galaxy properties do not clearly depend on cluster
mass for clusters more massive than $M_\ast\sim10^{14}M_\sun$, while
they correlate with cluster- centric distance.

    
 All these studies suggest that local processes, such as ram pressure,
strangulation and galaxy-galaxy interactions are the most easily detectable drivers of environmental evolution.
However, they must work simultaneously with processes
taking place on large scale, such as cluster-galaxy interactions, which apparently just leave more subtle signs.

\section*{Acknowledgments}
We thank the referee for their comments, which helped us to improve the manuscript.
Support for GLASS (HST-GO-13459) was provided by NASA through a grant
from the Space Telescope Science Institute, which is operated by the
Association of Universities for Research in Astronomy, Inc., under
NASA contract NAS 5-26555. We are very grateful to the staff of the
Space Telescope for their assistance in planning, scheduling and
executing the observations.  B.V. acknowledges the support from 
an  Australian Research Council Discovery Early Career Researcher Award (PD0028506).
\bibliographystyle{apj}
\bibliography{biblio_SFR2}

\end{document}